\documentclass[a4paper,11pt]{article}
\usepackage{pos}

\newcommand{\be}{\begin{equation}}
\newcommand{\ee}{\end{equation}}
\newcommand{\bea}{\begin{eqnarray}}
\newcommand{\eea}{\end{eqnarray}}
\newcommand{\beas}{\begin{eqnarray*}}
\newcommand{\eeas}{\end{eqnarray*}}

\newcommand{\la}{\langle}
\newcommand{\ra}{\rangle}
\newcommand{\corresp}{\hat{=}}
\newcommand{\Z}{\mathbb{Z}}
\newcommand{\R}{\mathbb{R}}

\title{Conjecture about the QCD phase diagram}

\author{Jos\'{e} Antonio Garc\'{\i}a-Hern\'{a}ndez}
\author{Edgar L\'{o}pez-Contreras}
\author{El\'{\i}as Natanael Polanco-Eu\'{a}n}
\author*{Wolfgang Bietenholz}

\affiliation{Instituto de Ciencias Nucleares,
  Universidad Nacional Aut\'{o}noma de M\'{e}xico \\
A.P.\ 70-543, C.P.\ 04510 Ciudad de M\'{e}xico, Mexico}

\emailAdd{joseantoniogarcia@ciencias.unam.mx}
\emailAdd{ed\underbar{~}lopez@ciencias.unam.mx}
\emailAdd{elias.polanco@correo.nucleares.unam.mx}
\emailAdd{wolbi@nucleares.unam.mx}

\abstract{We present a phase diagram study of the O(4) model as an
effective theory for 2-flavor QCD. In the chiral limit, both theories
perform spontaneous symmetry breaking with isomorphic groups,
which suggests that they belong to the same universality class.
Since we are interested in high temperature, we further assume dimensional
reduction to the 3d O(4) model, which implies topological sectors.
According to Skyrme and others, the corresponding topological charge
represents the baryon number. Hence the baryon chemical
potential $\mu_{B}$ appears as an imaginary vacuum angle, which can
be included in the lattice simulations without any sign problem.
We present simulation results for the critical line in the chiral
limit, and for the crossover in the presence of light quark masses.
The shapes of these lines are compatible with other conjectures, but
up to about $\mu_{B} \approx 300$~MeV we do not find a Critical Endpoint,
although there are indications for it to be near-by.}

\FullConference{%
 The 38th International Symposium on Lattice Field Theory, LATTICE2021
  26th-30th July, 2021
  Zoom/Gather@Massachusetts Institute of Technology
}

\begin{document}

\maketitle

\vspace*{-8mm}
\section{The O(4) model as an effective theory for 2-flavor QCD}
\vspace*{-2mm}

The QCD phase diagram is one of the most prominent open puzzles
within the Standard Model. What we know quite well is the behavior
at baryon chemical potential $\mu_{B}=0$, which corresponds to zero
baryon density. 

$\bullet$
For massless quarks $u$ and $d$, we expect a second order phase
transition. If we add the $s$-quark with its physical mass, the
critical temperature is around $T_{\rm c} \simeq 132~{\rm MeV}$
\cite{Ding}. For $2+1+1$ flavors, with $m_{u}=m_{d}=0$
and physical masses $m_{s}$ and $m_{c}$, the transition turns into
a crossover, but the pseudo-critical temperature is practically
the same, $T_{\rm pc} \simeq 134~{\rm MeV}$ \cite{Kotov}.

$\bullet$
  In the case of physical quark masses $m_{u}$, $m_{d}$, $m_{s}$,
  lattice studies reveal again a crossover, now at a pseudo-critical
  temperature of $T_{\rm pc} \simeq 155~{\rm MeV}$, see
  {\it e.g.}\ Refs.\ \cite{BazavovBhattacharya}.
  This value of $T_{\rm pc}$ is compatible with the
  experimentally measured freeze-out temperature
  of the quark-gluon plasma.

At $\mu_{B} > 0$, lattice QCD is plagued by a notorious
``sign problem'' (see {\it e.g.}\ Ref.\ \cite{Philippe}),
which prevents the exploration of the QCD phase
diagram at finite baryon density. Despite numerous
efforts, there is still no breakthrough in the attempts to overcome
this sign problem. Therefore it is still appropriate to elaborate
conjectures about the QCD phase diagram based on effective theories.

Here we consider the O(4) non-linear $\sigma$-model, which is assumed
to be in the same universality class as 2-flavor QCD in the chiral
limit \cite{chiralPT}.
Adding an external ``magnetic field'' (or ``ordering field'') $\vec h$
corresponds to a degenerate quark mass, and we obtain the Euclidean action
\be
S [ \vec e \, ] = \int d^{3}x \int_{0}^{\beta} dx_{4} \
\Big[ \frac{F_{\pi}^{2}}{2}
  \partial_{\mu} \vec e (x) \cdot \partial_{\mu} \vec e (x)
  - \vec h \cdot \vec e (x) \Big] \ , \quad \vec e(x) \in S^{3}
\ , \ F_{\pi} \simeq 92.4 \ {\rm MeV} \ .
\ee
At $h = |\vec h | = 0$ the action has a global O(4) symmetry,
which can break spontaneously to O(3). $h >0$ adds some explicit
symmetry breaking, in analogy to degenerate quark masses
$m_{u} = m_{d} > 0$.
There is a local isomorphy with or without symmetry breaking,
$$
\{ \ {\rm SU}(2)_{\rm L} \otimes {\rm SU}(2)_{\rm R} \ \corresp \
   {\rm O}(4) \ \} \quad \longrightarrow \quad 
\{ \ {\rm SU}(2)_{\rm L=R} \ \corresp \ {\rm O}(3) \ \} \ ,
$$
which supports the assumption that the systems near
criticality are in the same universality class.

We are interested in high temperature $T = 1/\beta$, so we assume
dimensional reduction to the 3d O(4) model,
\be
S [\vec e \, ] = \beta \int d^{3}x \
\Big[ \frac{F_{\pi}^{2}}{2}
  \partial_{i} \vec e (x) \cdot \partial_{i} \vec e (x)
  - \vec h \cdot \vec e (x) \Big] = \beta H[ \vec e \, ] \ .
\ee
We are going to consider cubic lattice volumes, $V = L^{3}$.
With periodic boundary conditions, the configurations
fall in topological sectors, $\pi_{3}(S^{3}) = \Z$. As pointed out by
Skyrme and others, the topological charge $Q$ corresponds to the
baryon number $B$ \cite{Skyrme}. The field $\vec e(x)$ seems to represent
just the pions, but its topological windings also account for baryons.
Hence the baryon chemical potential $\mu_{B}$ corresponds to an imaginary
vacuum angle $\theta$, which extends the Hamilton function to
$H[\vec e \, ] = \dots - \mu_{B} Q[\vec e \,] \in \R$.

\vspace*{-2mm}
\section{Quark mass and chemical potential without sign problem}
\vspace*{-2mm}

\begin{figure}[h!]
\vspace*{-4mm}
\begin{center}
\includegraphics[angle=0,width=.28\linewidth]{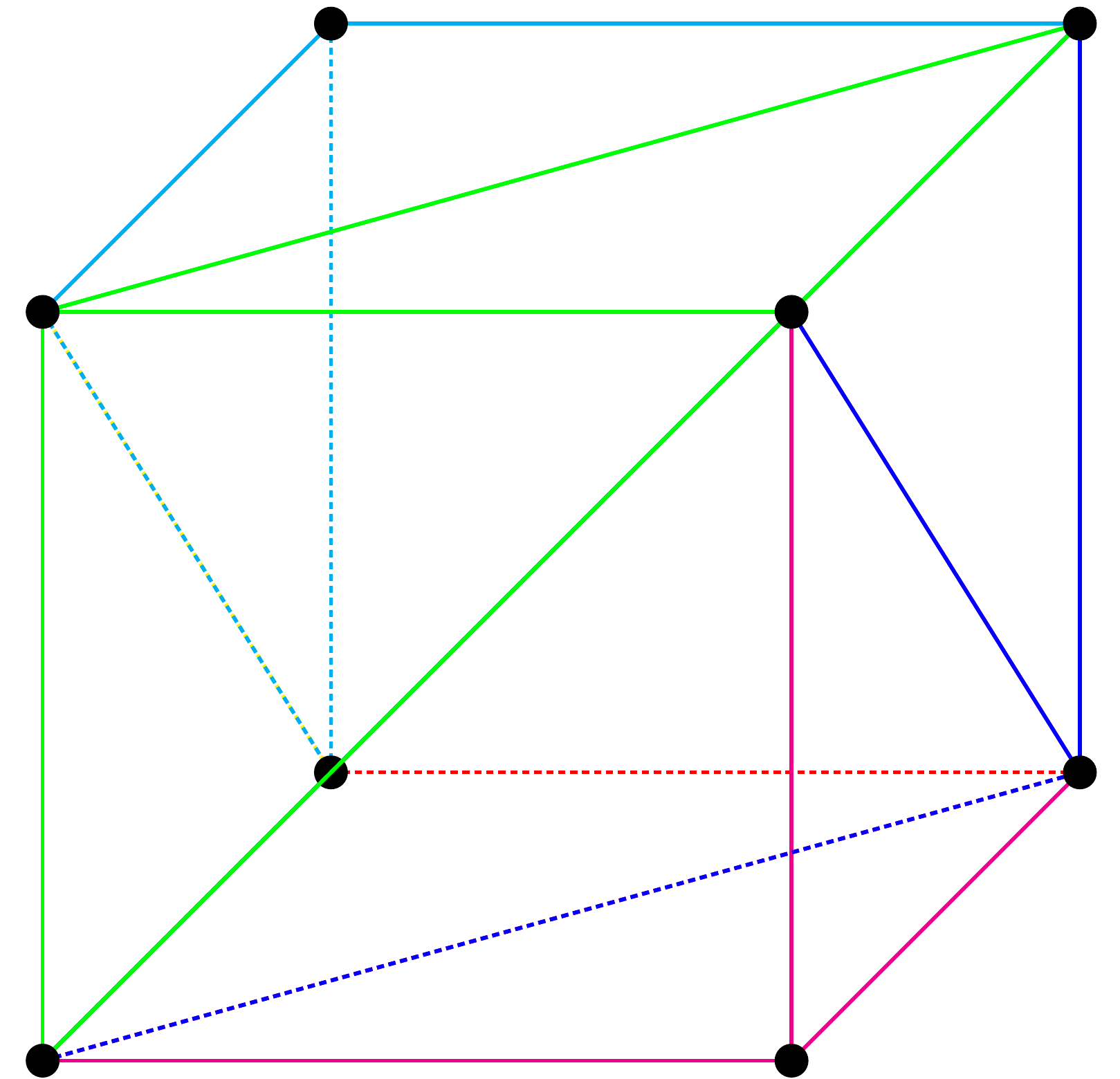}
\hspace*{2cm}
\includegraphics[angle=0,width=.3\linewidth]{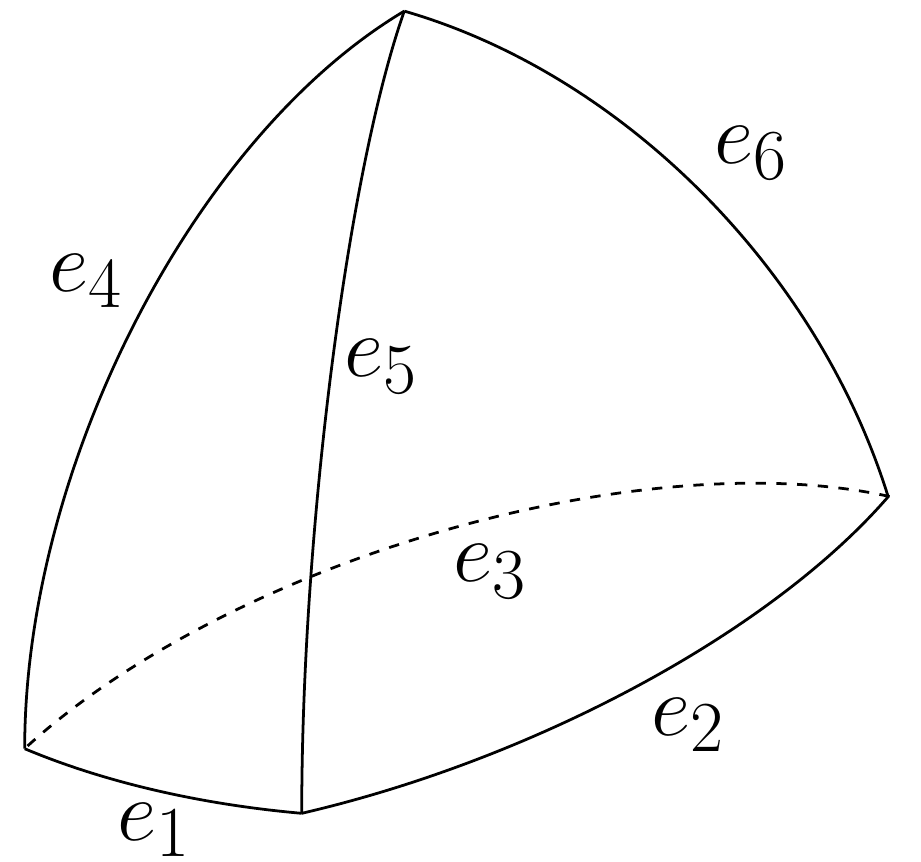}
\end{center}
\vspace*{-4mm}
\caption{Left: division of a lattice unit cube into 6 tetrahedra.
  Right: symbolic sketch of a spherical tetrahedron on $S^{3}$. The
  topological charge is composed of the oriented volumes of such
  spherical tetrahedra.}
\vspace*{-1mm}
\label{Qgeom}
\end{figure}

We are going to use the standard lattice action (in lattice units)
\be
S_{\rm lat}[\vec e \, ] = -\beta_{\rm lat} \Big( \sum_{\la xy \ra}
\vec e_{x} \cdot \vec e_{y} + \vec h_{\rm lat} \cdot \sum_{x} \vec e_{x}
+ \mu_{B,{\rm lat}} Q[\vec e \,] \Big) \ ,
\ee
with $\beta_{\rm lat} = \beta F_{\pi}^{2}$, $\vec h_{\rm lat} = \vec h / F_{\pi}^{2}$,
$\mu_{B, {\rm lat}} =  \mu_{B} / F_{\pi}^{2}$.
For the topological charge of a lattice configuration, we employ the
geometric definition, which assures $Q[\vec e \,] \in \Z$.
We split each lattice unit cube into 6 tetrahedra, as illustrated in
Figure \ref{Qgeom} (left). The 4 spins attached to the vertices of
one tetrahedron, say $(\vec e_{w}, \vec e_{x}, \vec e_{y}, \vec e_{z})$,
span a {\em spherical tetrahedron} on $S^{3}$, as symbolically sketched
in Figure \ref{Qgeom} (right): the edges $e_{1} \dots e_{6}$ are
geodesics in $S^{3}$. The topological density of a tetrahedron is
given by the oriented volume of this spherical tetrahedron,
$V_{w,x,y,z}[\vec e \,] \in (-\pi^{2}, \pi^{2})$, which is normalized
such that
\be
Q[\vec e\,] = \frac{1}{2\pi^{2}} \sum_{\la wxyz \ra} V_{w,x,y,z}[\vec e\,]
\in \Z \ .
\ee

We have implemented a rather involved, implicit formula for
$V_{w,x,y,z}[\vec e\,]$, which was derived in Ref.\ \cite{Murakami}.
However, it is computationally more efficient to choose some
reference point on $S^{3}$ and count the tetrahedra which
enclose this point, in an oriented sense.
(We checked extensively that the results coincide.)

\begin{figure}
\vspace*{-7mm}
\begin{center}
\includegraphics[angle=0,width=.42\linewidth]{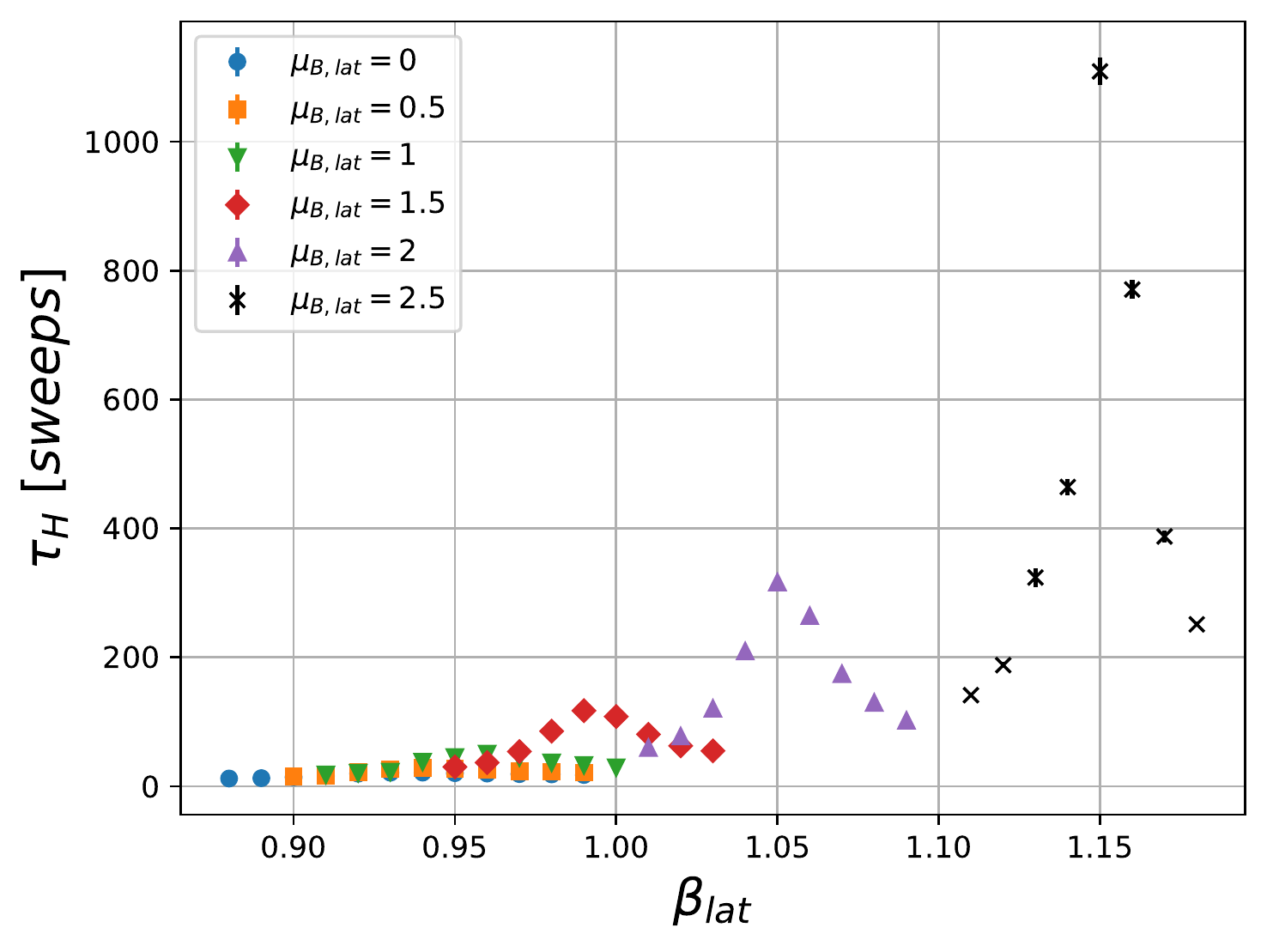}
  \hspace*{5mm}
\includegraphics[angle=0,width=.42\linewidth]{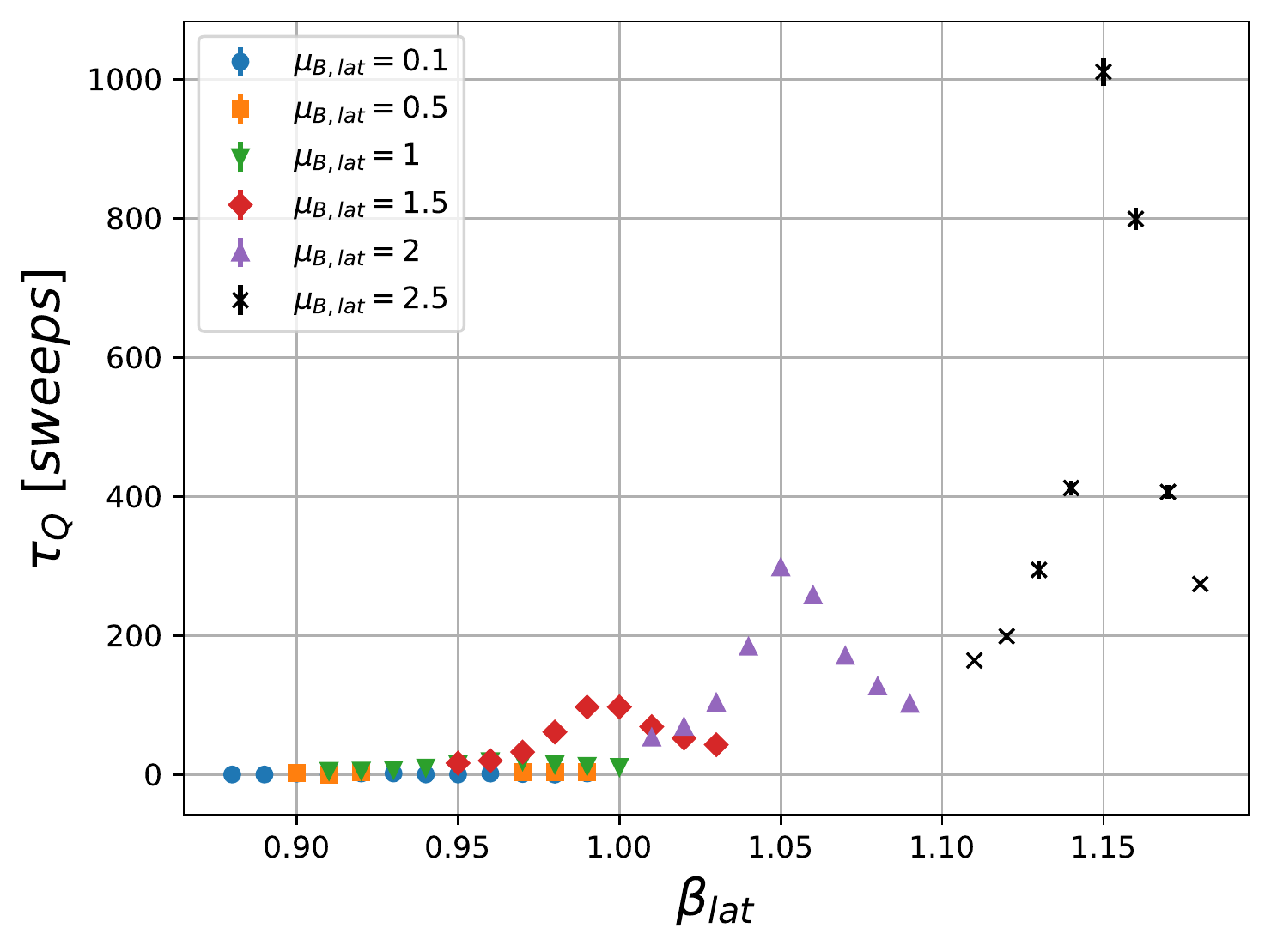}
\end{center}
\vspace*{-6mm}
\caption{The auto-correlation time $\tau$, in units of multi-cluster
  update sweeps, for the energy ($\tau_{H}$) and the topological charge
 ($\tau_{Q}$), which are very similar. These examples refer to $L=20$, $h=0$.}
\label{autocor}
\vspace*{-4mm}
\end{figure}

The applicability of an efficient cluster algorithm is another
benefit of the O(4) model as an effective theory. We used the
multi-cluster Wolff algorithm, which is particularly useful to
avoid topological freezing. Different values of $\mu_{B}$ are taken
into account by modifying the cluster flip probability, along the
lines of Ref.\ \cite{Wang}.
As an example, Figure \ref{autocor} shows the exponential auto-correlation
time for the energy $H$, $\tau_{H}$, and for the topological charge $Q$,
 $\tau_{Q}$, in the lattice volume $V = 20^{3}$, at $h=0$. They
hardly differ, thanks to the cluster algorithm. Still, for increasing
$\mu_{B,{\rm lat}}$, $\tau$ rises rapidly: at $\mu_{B,{\rm lat}} = 2.5$ it
already exceeds 1000 multi-cluster sweeps, which has prevented
us (so far) from exploring $\mu_{B,{\rm lat}} > 2.5$. In all our data sets,
the measurements were separated at least by $2 \tau$; without being very
careful, one can indeed be led to wrong conclusions about a Critical
Endpoint (CEP). From Figure \ref{autocor} we already suspect that
we stay with a second order phase transition up to $\mu_{B,{\rm lat}} = 2.5$,
where $\tau$ would diverge in infinite volume at the critical
value $\beta_{\rm c,lat}$.

\section{Conjectured phase diagram in the chiral limit}
\label{sec:chiral}

We first present our results at $h=0$, {\it i.e.}\ in the chiral limit,
where we deal with a second order phase transition, at least at low
$\mu_{B,{\rm lat}}$. We convert lattice units to physical units by referring
to the critical temperature at $\mu_{B}=0$. On the lattice, it was
measured to high precision, $\beta_{\rm c,lat}=0.93590$ \cite{Oevers},
which we match with the aforementioned lattice QCD
result of $T_{\rm c} \simeq 132 \, {\rm MeV}$, such that
\be
\mu_{B} = \frac{\beta_{\rm c,lat}}{\beta_{\rm c}} \mu_{B,{\rm lat}}
\approx 124 \ {\rm MeV} \ \mu_{B,{\rm lat}} \ .
\ee
We simulated at $\mu_{B,{\rm lat}} = 0,\ 0.1,\ 0.2,\ \dots 1.5, \ 2,\ 2.5$,
which corresponds to the range $\mu_{B} = 0 \dots 309 \ {\rm MeV}$.
We are restricted to modest lattice volumes $V=L^{3}$,
$L = 10, \ 12,\ 16,\ 20$, due to the problem with the huge
auto-correlation time at our largest values of $\mu_{B,{\rm lat}}$.
For each parameter set, we performed $10^{4}$ measurements, with
perfectly thermalized and de-correlated configurations.
In order to monitor the phase
transition and its order, we measured some observables given by first
and second derivatives of the free energy $F = - T \ln Z$ \cite{Edgar}.

\begin{figure}
\vspace*{-8mm}
\begin{center}
\includegraphics[angle=0,width=.44\linewidth]{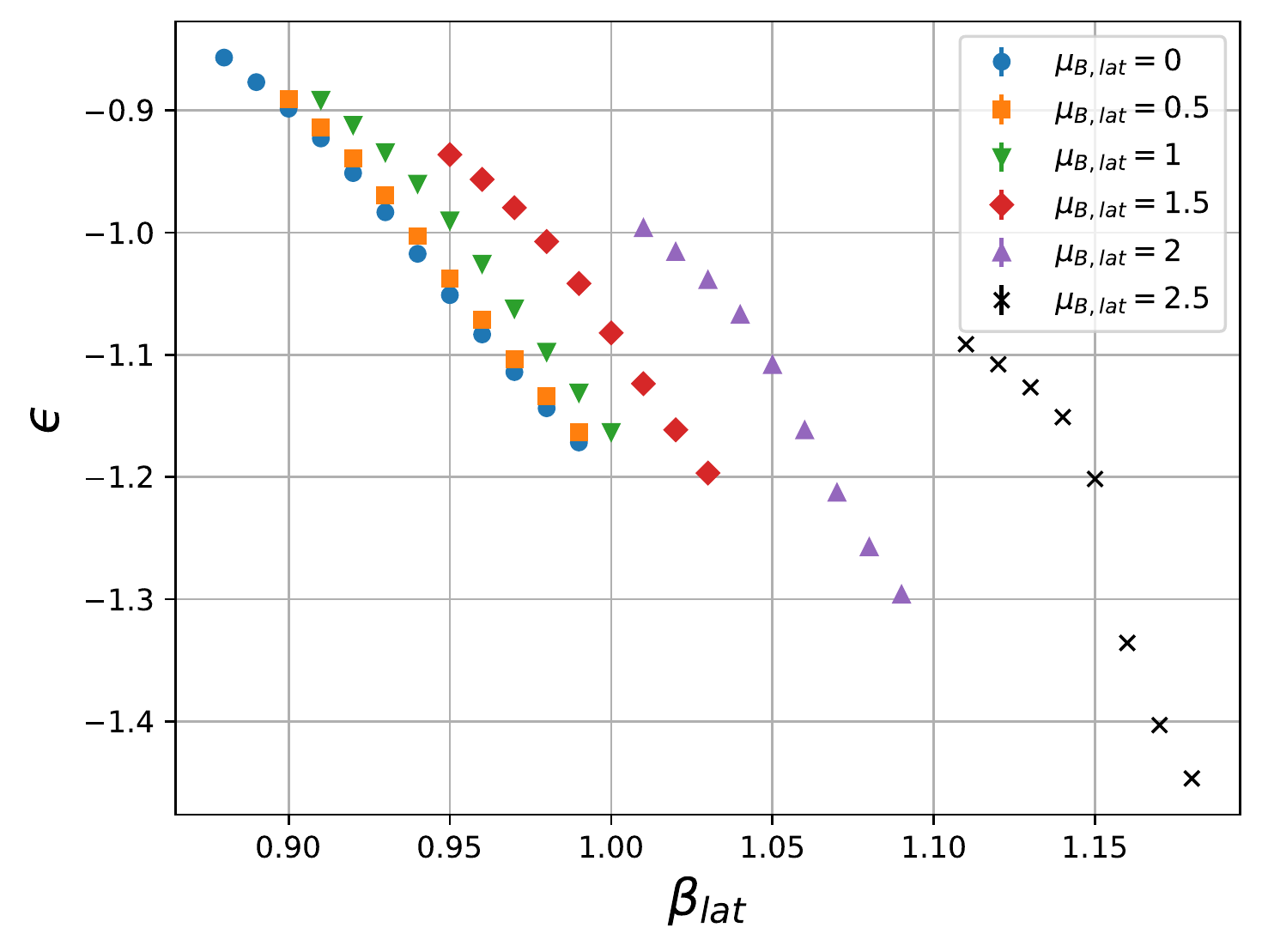}
\hspace*{10mm}
\includegraphics[angle=0,width=.44\linewidth]{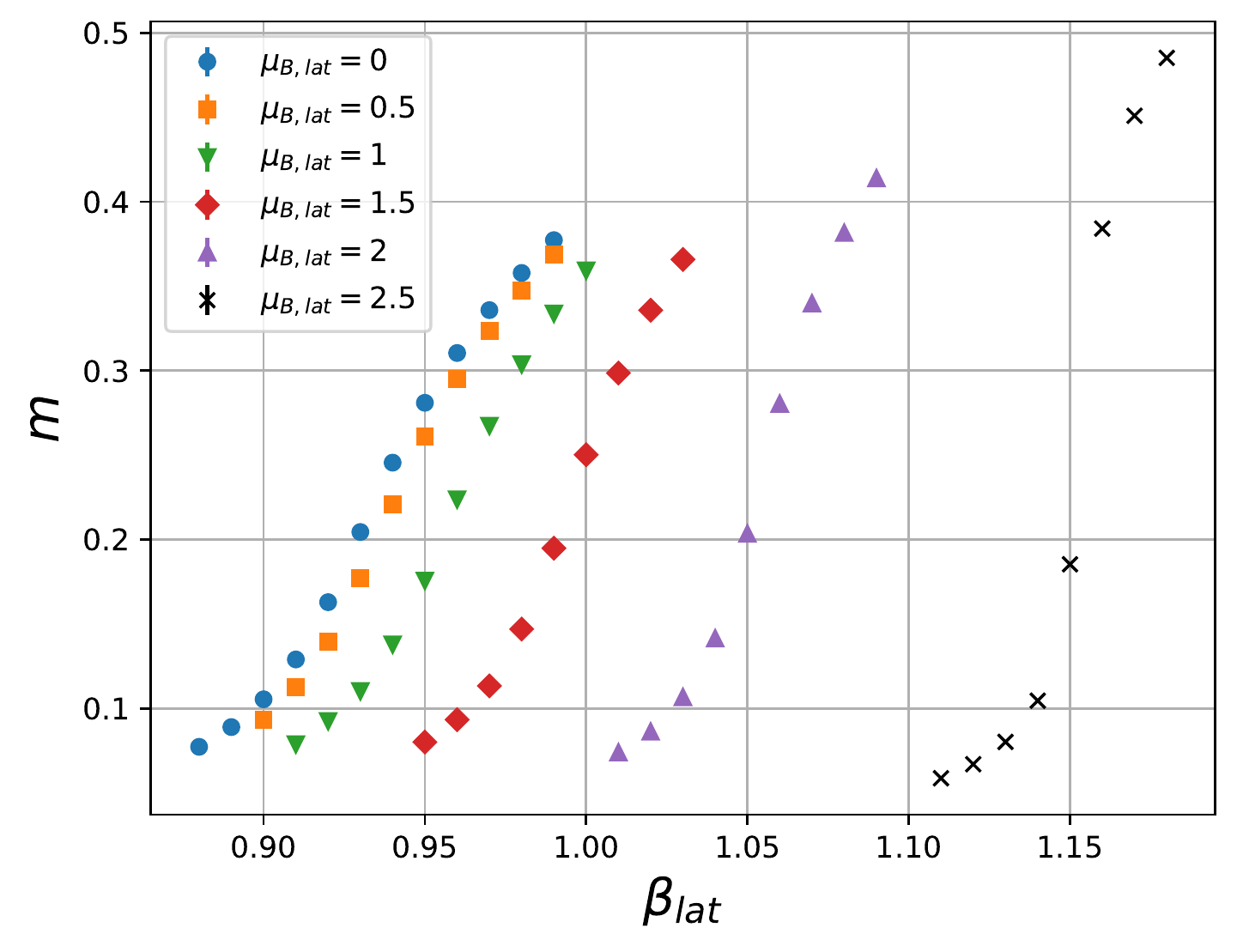} \\
\includegraphics[angle=0,width=.44\linewidth]{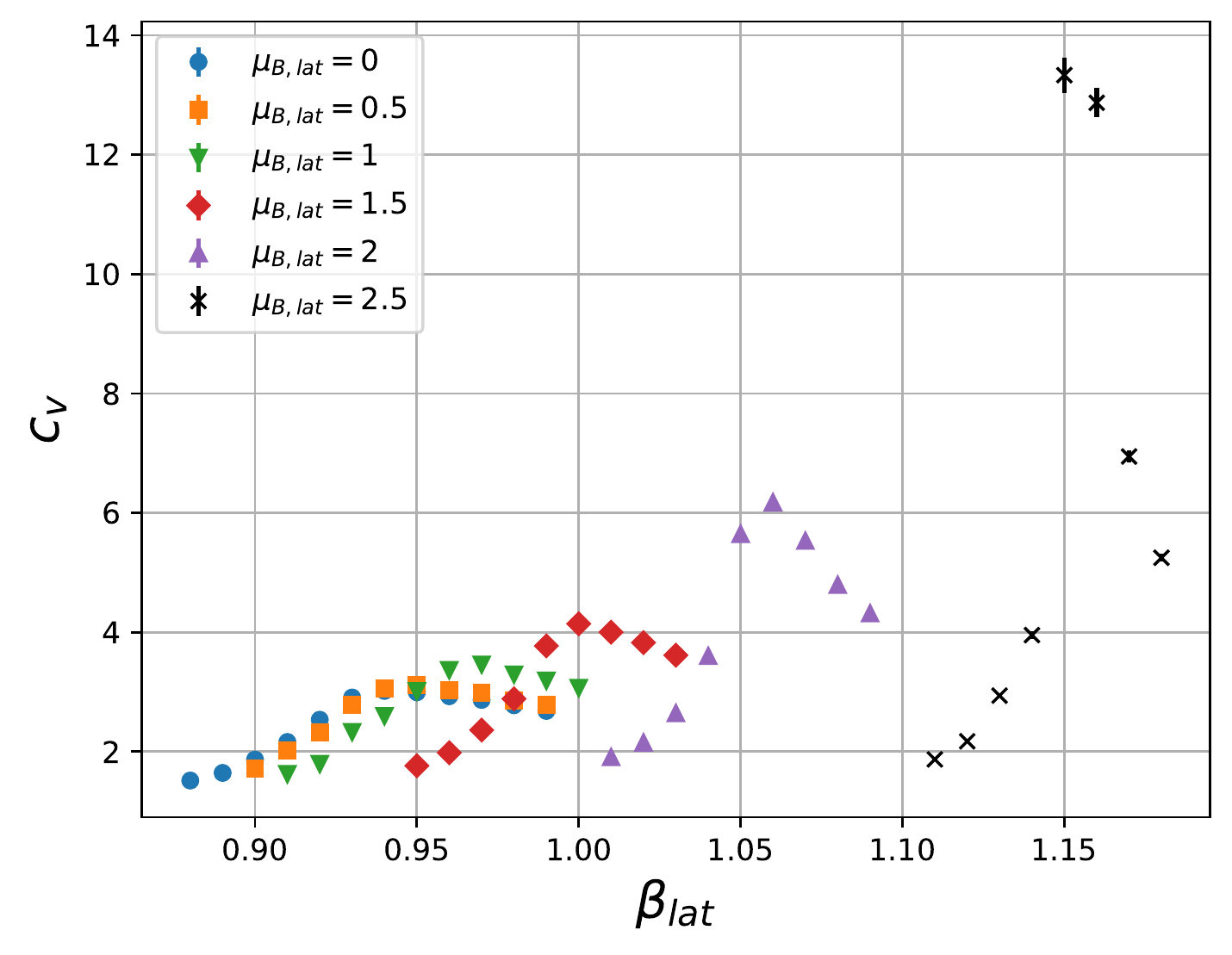}
\hspace*{10mm}
\includegraphics[angle=0,width=.44\linewidth]{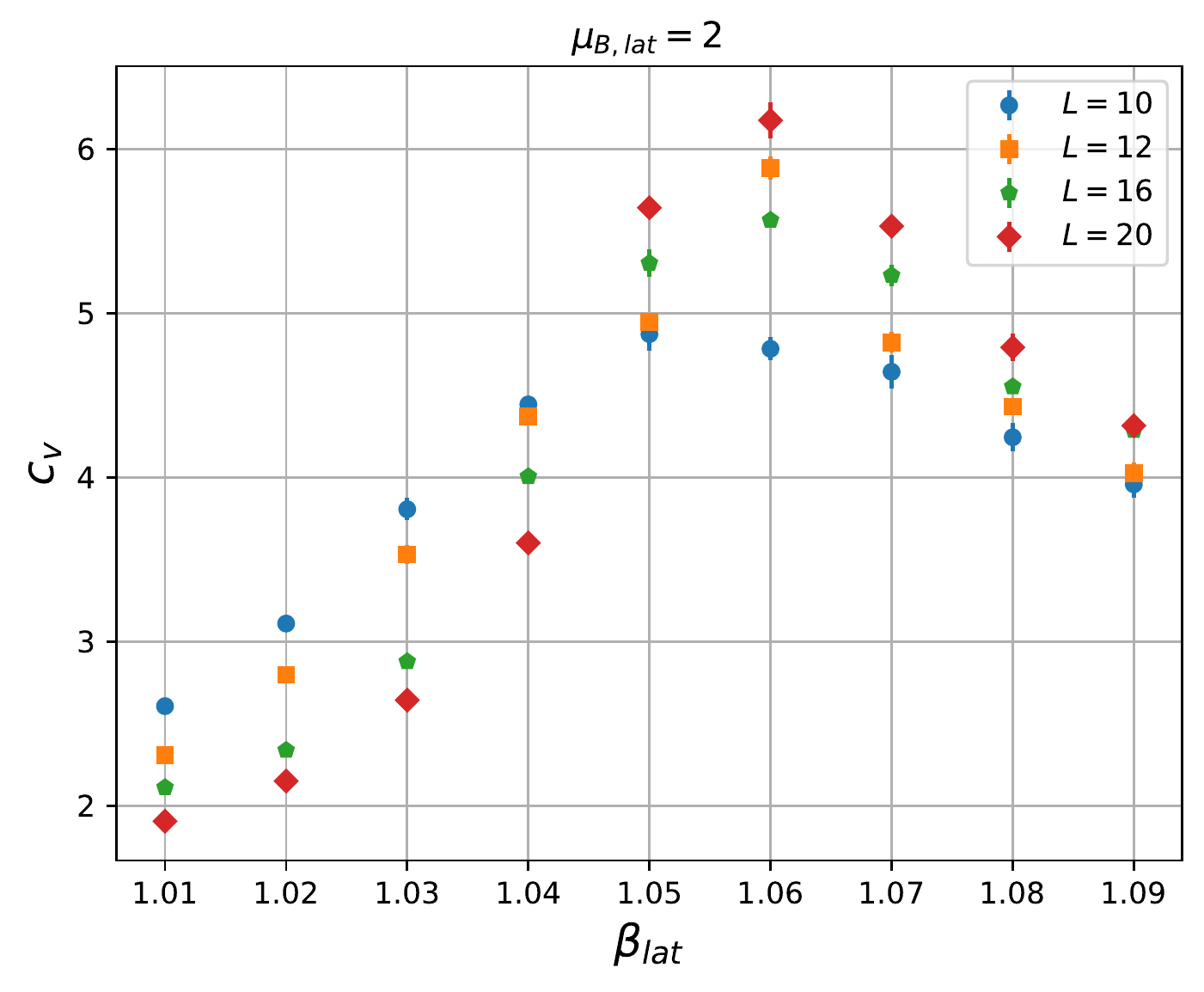} \\
\includegraphics[angle=0,width=.44\linewidth]{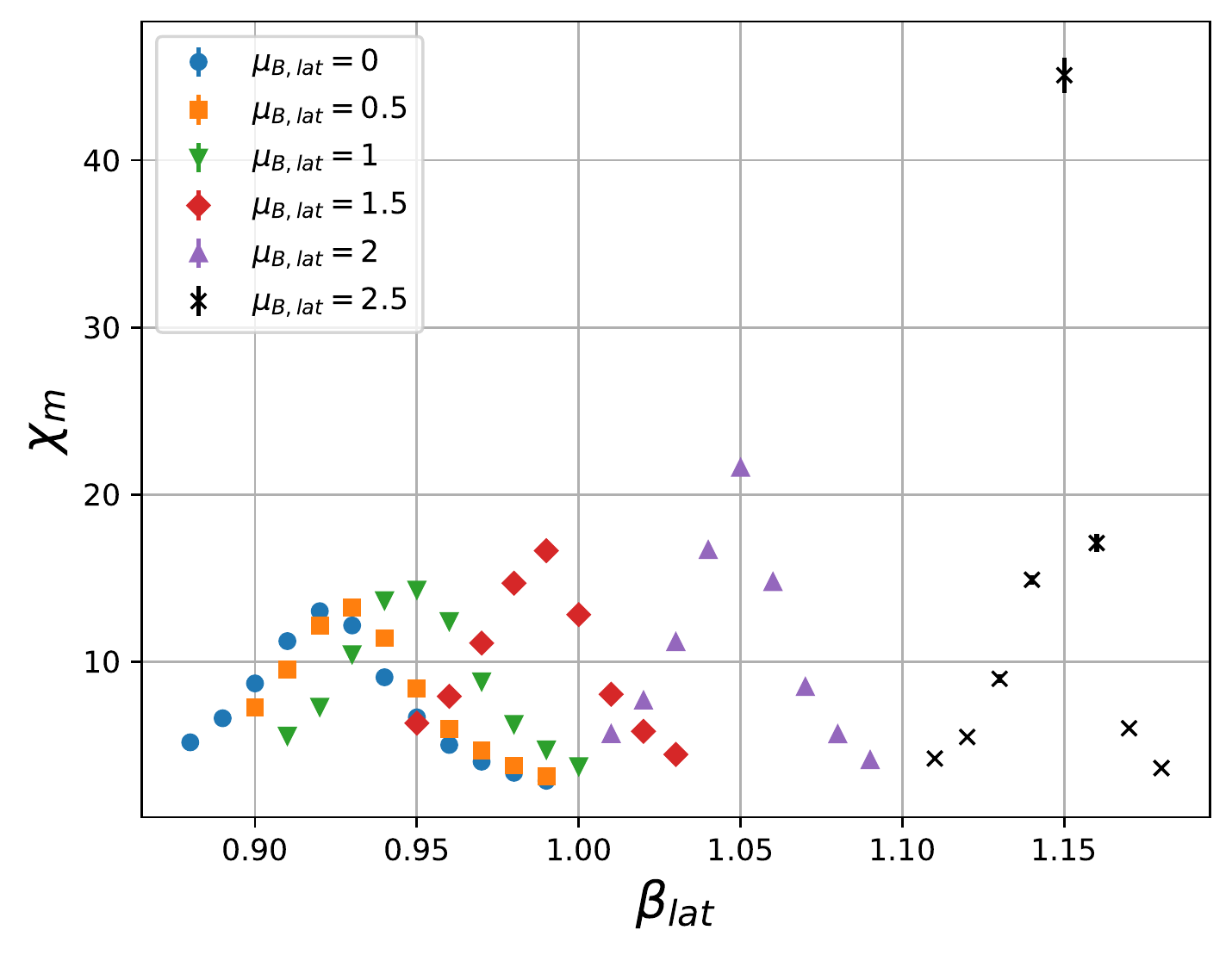}
\hspace*{10mm}
\includegraphics[angle=0,width=.44\linewidth]{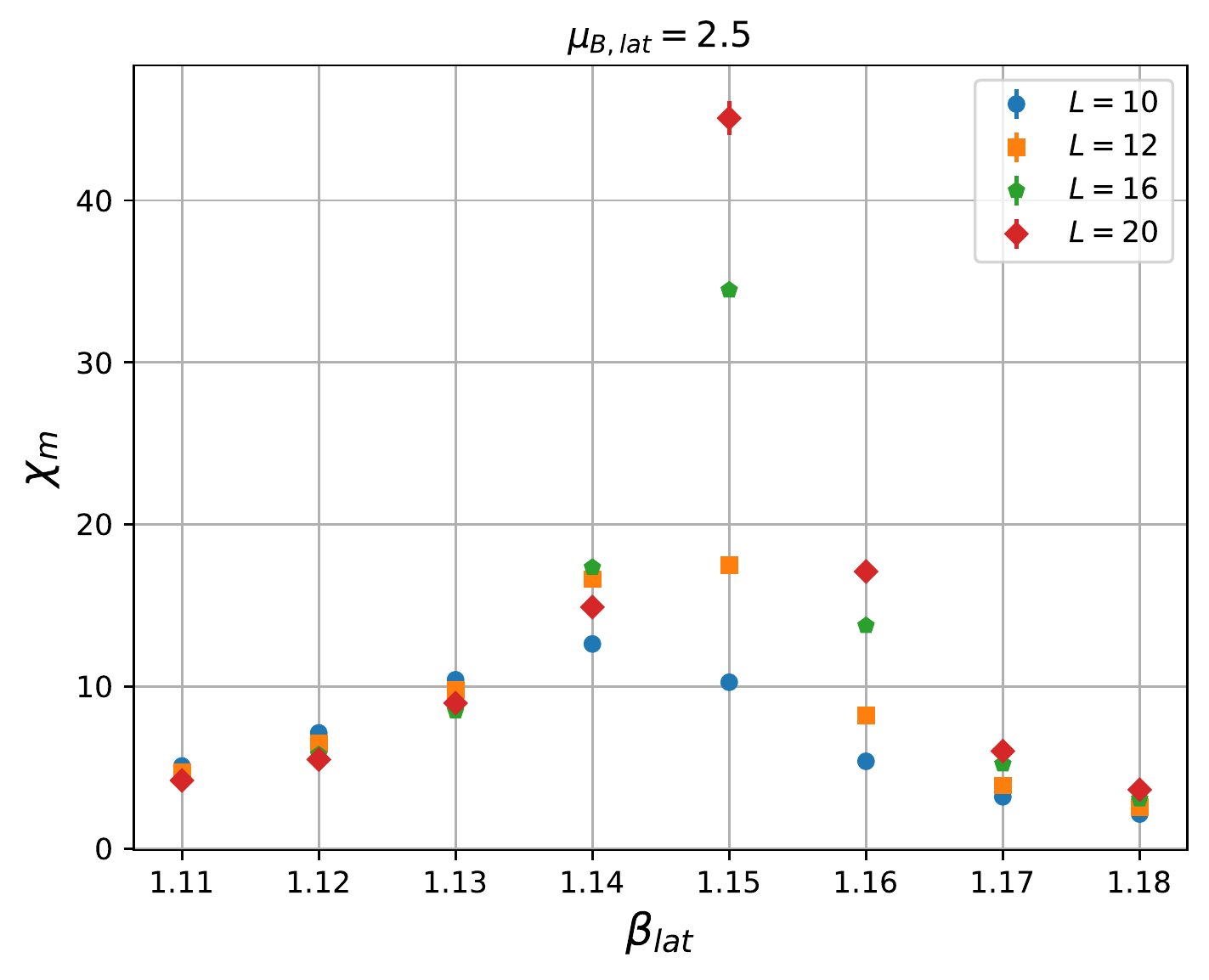}
\end{center}
\vspace*{-6mm}
\caption{Simulation results at $h=0$: energy density $\epsilon$ and
  magnetic density $m$ (top), specific heat $c_{V}$ and magnetic
  susceptibility $\chi_{m}$ (center and bottom). The $\beta_{\rm lat}$
  interval where $\epsilon$ and $m$ have maximal slopes agrees with the
  peaks of $c_{V}$ and $\chi_{m}$. We also show how the peak heights
  increase with the volume.}
\label{h0multiplots}
\end{figure}

The plots in Figure \ref{h0multiplots} on top show the energy
density $\epsilon = \la H \ra /V$ and the magnetization density
$m = \la |\vec M | \ra /V$, $\vec M = \sum_{x} \vec e_{x}$ (the
order parameter), which are first derivatives of $F$, at $L=20$.
We see that increasing $\mu_{B,{\rm lat}}$
at fixed $\beta_{\rm lat}$ enhances $\epsilon$ but reduces $m$,
as expected when more topological windings appear.
In particular for our largest values of $\mu_{B,{\rm lat}}$ we clearly
see intervals of maximal slope, which coincide and move to larger
$\beta_{\rm lat}$ when $\mu_{B,{\rm lat}}$ increases. This indicates the
approximate value of $\beta_{\rm c,lat}$. At $\mu_{B,{\rm lat}} = 2.5$ these
slopes turn into quasi-jumps, so one might wonder whether a first
order phase transition is near-by, or even attained.

The lower four plots in Figure \ref{h0multiplots} refer to second
derivatives of $F$, namely the specific heat $c_{V}$ and the magnetic
susceptibility $\chi_{m}$,
\vspace*{-2mm}
\be
c_{V} = \frac{\beta^{2}_{\rm lat}}{V} \Big( \la H^{2} \ra - \la H \ra^{2} \Big)
\ , \quad
\chi_{m} = \frac{\beta_{\rm lat}}{V} \Big( \la \vec M^{\, 2} \ra -
\la \vec M \ra^{2} \Big) \ .
\vspace*{-1.5mm}
\ee
At $L=20$ (plots on the left), we see pronounced peaks, in particular
at $\mu_{B,{\rm lat}}=2$ and $2.5$, hence the corresponding phase transition
(in infinite volume) is likely to be still second order. Regarding the
$L$-dependence at fixed $\mu_{B,{\rm lat}}$ (plots on the right), the peak
location of $c_{V}$ hardly moves, but in the case of $\chi_{m}$ a
thermodynamic extrapolation is important; it leads to values of
$\beta_{\rm c,lat}$, which are consistent with $c_{V}$ and other criteria.

For a second order phase transition, the peak height of $c_{V}$ is
$\propto L^{\alpha / \nu}$; in the example of Figure \ref{h0multiplots}
we obtain for the ratio of critical exponents $\alpha / \nu \approx 0.2$.
The peak height of $\chi_{m}$ has a stronger $L$-dependence:
in the range $\mu_{B,{\rm lat}}=0 \dots 1.5$ we obtain
$\gamma /\nu = 1.9(2)$, which is compatible with the
precise result at $\mu_{B,{\rm lat}}=0$ from Ref.\ \cite{Oevers},
$\gamma /\nu = 1.970$.
Taking all that together strongly supports second order up to
 $\mu_{B,{\rm lat}}=2.5$.
\begin{figure}
\vspace*{-1mm}
\begin{center}
\includegraphics[angle=0,width=.44\linewidth]{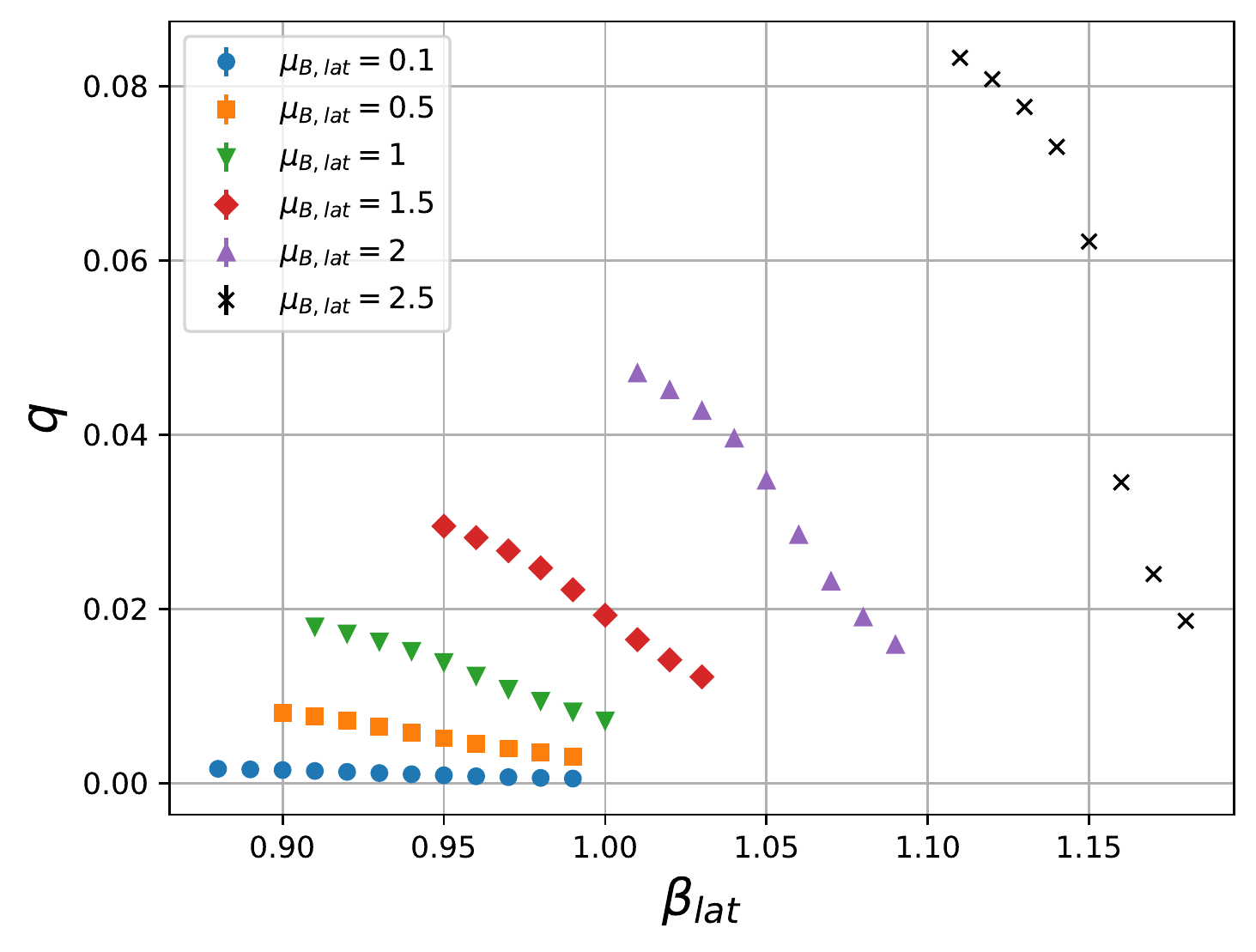}
\hspace*{10mm}  
\includegraphics[angle=0,width=.44\linewidth]{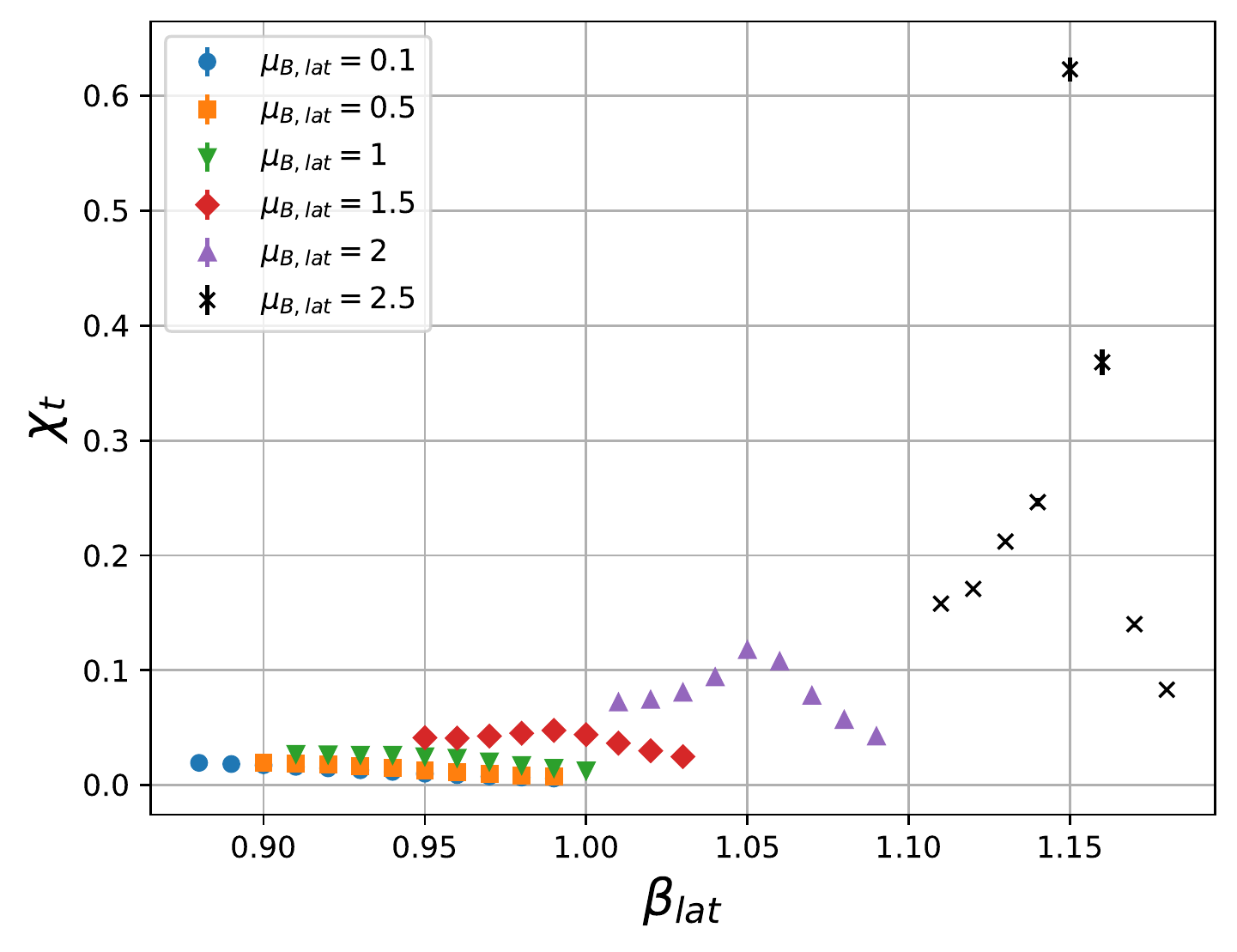}
\end{center}
\vspace*{-6mm}
\caption{Topological charge density $q = \la Q\ra/V$ and the
topological susceptibility $\chi_{\rm t}$, in the chiral limit.}
\label{h0Q}
\vspace*{-3mm}
\end{figure}

Figure \ref{h0Q} (left) shows the topological charge density
$q = \la Q \ra /V = \partial_{\mu_{B,\rm lat}} F$, which vanishes at
$\mu_{B,{\rm lat}}=0$ due to parity invariance. Topological winding is
enhanced by $\mu_{B,{\rm lat}} > 0$, but suppressed by low temperature.
The plot on the right shows the topological susceptibility
$\chi_{\rm t} = ( \la Q^{2} \ra - \la Q \ra^{2} )/V$. Again we see
pronounced peaks for $\mu_{B,{\rm lat}} \geq 1.5$, at locations which
agree with the previous determinations of $\beta_{\rm c}$. If we
use the peak height to define a ratio of critical exponents, in analogy
to $c_{V}$ and $\chi_{m}$, $\chi_{\rm t} (\beta_{\rm c,lat}) \propto
L^{x/\nu}$, we obtain {\it e.g.}\
$x/\nu |_{\mu_{B,{\rm lat}=0}} \simeq 0.2$,\,
$x/\nu |_{\mu_{B,{\rm lat}=1}} \simeq 0.3$.

\begin{figure}
\vspace*{-4mm}
\begin{center}
\includegraphics[angle=0,width=.55\linewidth]{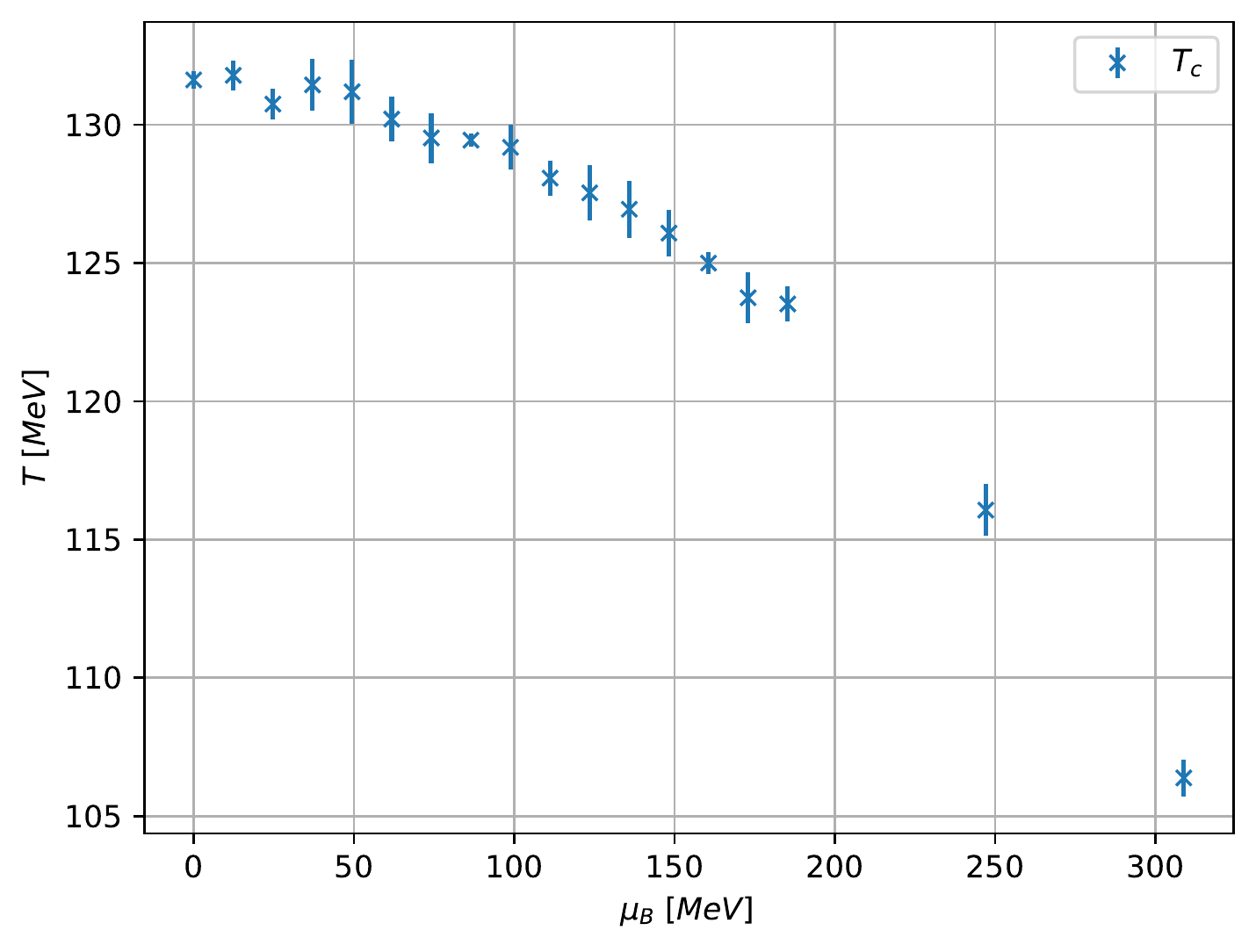}
\end{center}
\vspace*{-5mm}
\caption{Conjectured phase diagram for QCD with 2 massless quark flavors.}
\vspace*{-3mm}
\label{chiralphasediagram}
\end{figure}
Finally we combine all the determinations of $\beta_{\rm lat}(\mu_{B,{\rm lat}})$
(based on steepest slopes and peaks, also considering the correlation
length, all extrapolated to $V \to \infty$),
and convert to physical units. This leads to the phase diagram in the
chiral limit shown in Figure \ref{chiralphasediagram}. $T_{\rm c}$ decreases
for increasing $\mu_{B}$, in agreement with the consensus in the
literature, but we do not observe a CEP ({\it i.e.}\ a change to a first
order phase transition) in the regime $\mu_{B} \lesssim 309 \ {\rm MeV}$
and $T \gtrsim 106 \ {\rm MeV}$.

\vspace*{-2mm}
\section{Conjectured phase diagram with light quarks}
\vspace*{-2mm}

We proceed to a finite, degenerate quark mass $m_{\rm q}$ of the light
flavors $u$ and $d$ \cite{JAGH}, which corresponds to the external
magnetic field strength $h = |\vec h|$. In lattice units we choose the
values $h_{\rm lat}=0.367$ and $0.14$. Here we refer to the
pseudo-critical crossover temperature
$T_{\rm pc} \simeq 155 \ {\rm MeV}$
to relate lattice units and physical units,
$h \approx h_{\rm lat} \, T_{\rm pc}^{4} / T_{\rm pc,lat}^{4}$.
Our results for $T_{\rm pc,lat} = 1/\beta_{\rm pc,lat}$ are ambiguous,
as we see from the maxima of $c_{V}$ and $\chi_{m}$ (see below). If we take
the average, we obtain $T_{\rm pc,lat} (h_{\rm lat} =0.367) \simeq 1.273$ and
$T_{\rm pc,lat} (h_{\rm lat}=0.14) \simeq 1.172$, in a monotonous relation to
the chiral limit, $T_{\rm c,lat}(h_{\rm lat}=0) = 1.06849$ \cite{Oevers}.
For a dimensioned quantity like $\mu_{B}$ this suggests 
\be
\mu_{B} \approx 122~{\rm MeV} \, \mu_{B,{\rm lat}} \quad (h_{\rm lat} = 0.367)
\ ; \qquad
\mu_{B} \approx 132~{\rm MeV} \, \mu_{B,{\rm lat}} \quad (h_{\rm lat} = 0.14) \ .
\ee
If we interpret $h$ as $m_{\rm q} \Sigma$, and insert for the chiral
condensate $\Sigma \simeq (250~{\rm MeV})^{3}$, we obtain the quark mass
$m_{\rm q} \approx 5.2~{\rm MeV}$ and $m_{\rm q} \approx 2.7~{\rm MeV}$,
which is compatible with $m_{d}$ and with $m_{u}$, respectively.

We incorporate the additional term by again adjusting the cluster
flip probability, following Ref.\ \cite{Wang}. We also implemented
an alternative method, which adds a global ``ghost field'' with
possible bonds to any spin variable \cite{ForKas}.

\begin{figure}[h!]
\vspace*{-5mm}
\begin{center}
\includegraphics[angle=0,width=.36\linewidth]{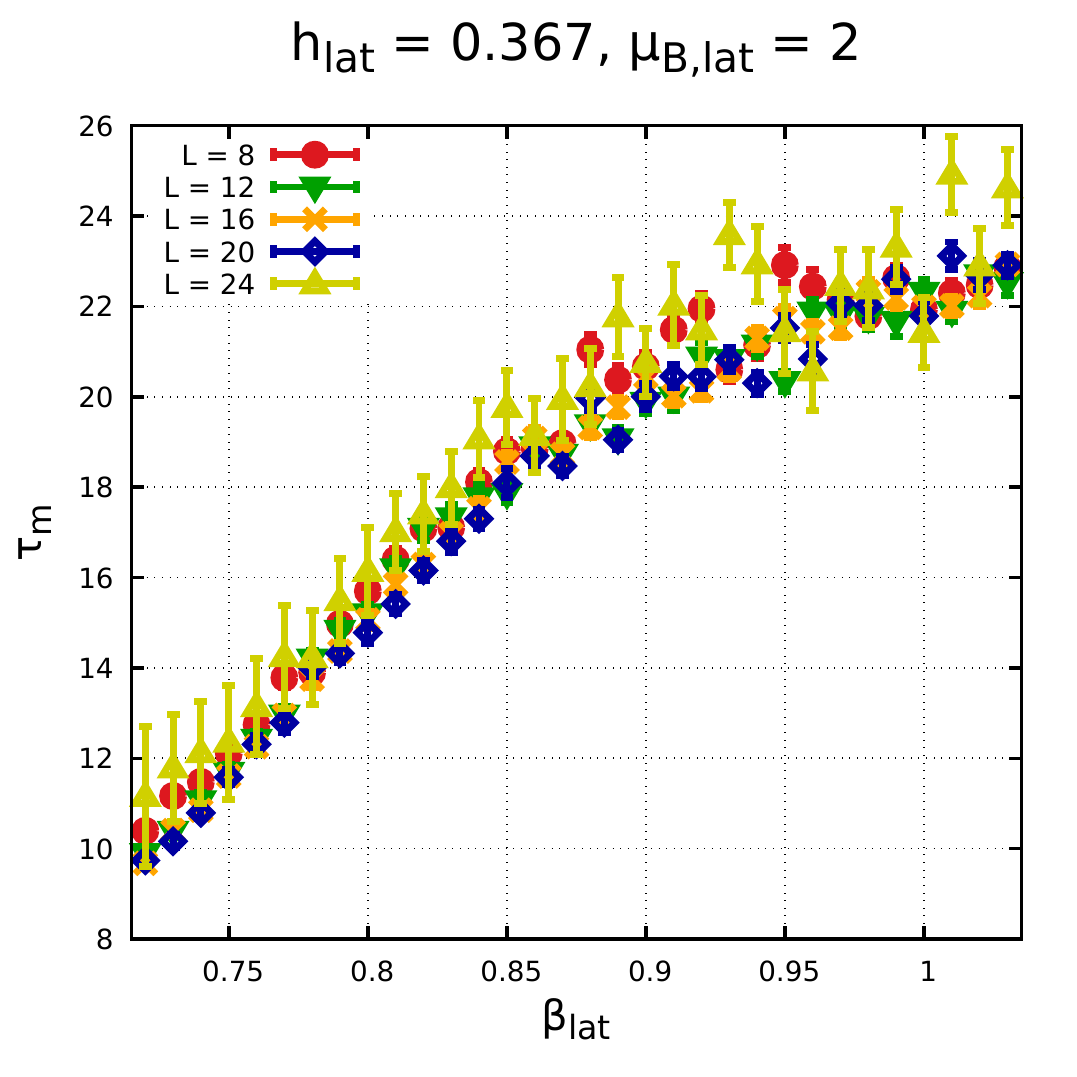}
\hspace*{10mm}
\includegraphics[angle=0,width=.36\linewidth]{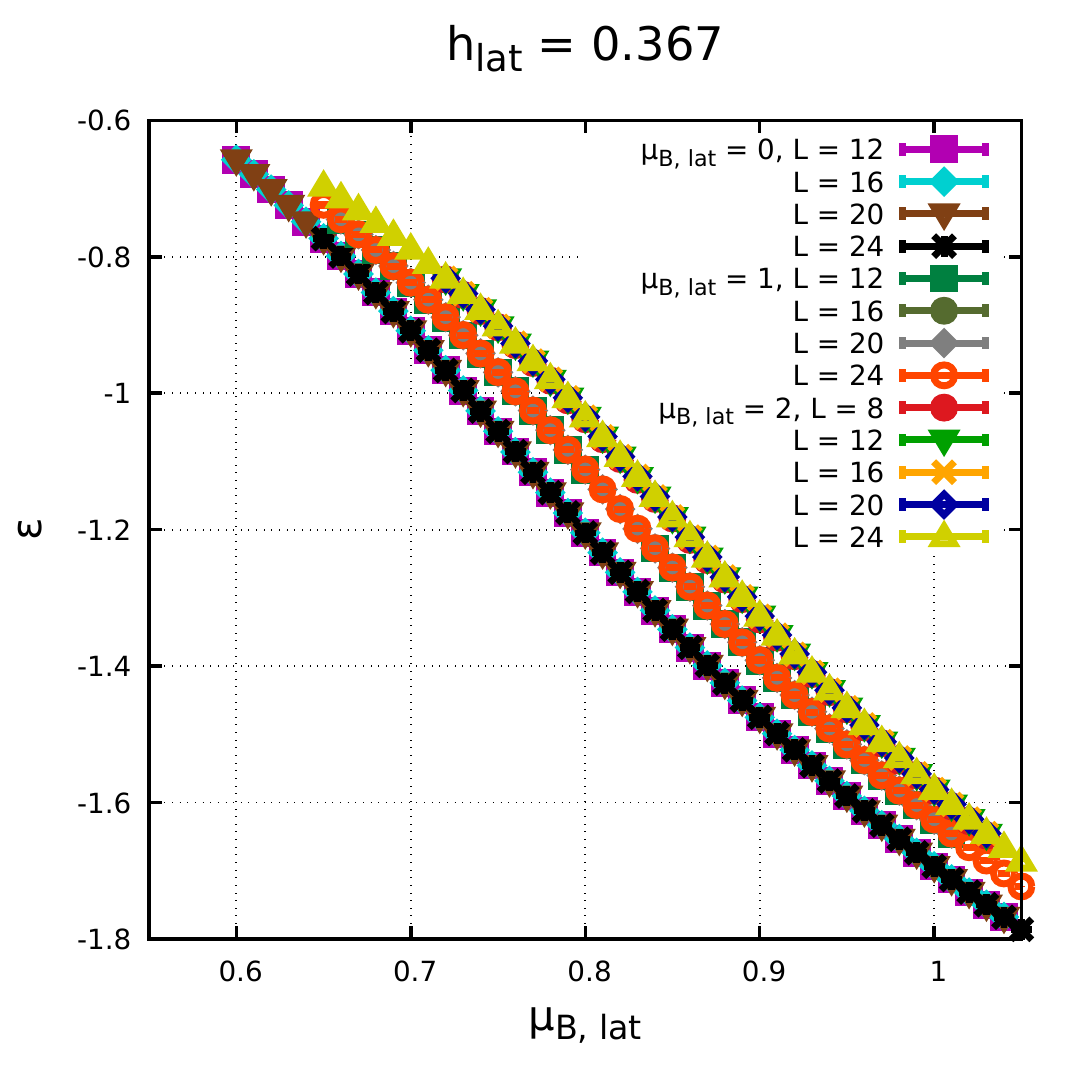} \\
\includegraphics[angle=0,width=.36\linewidth]{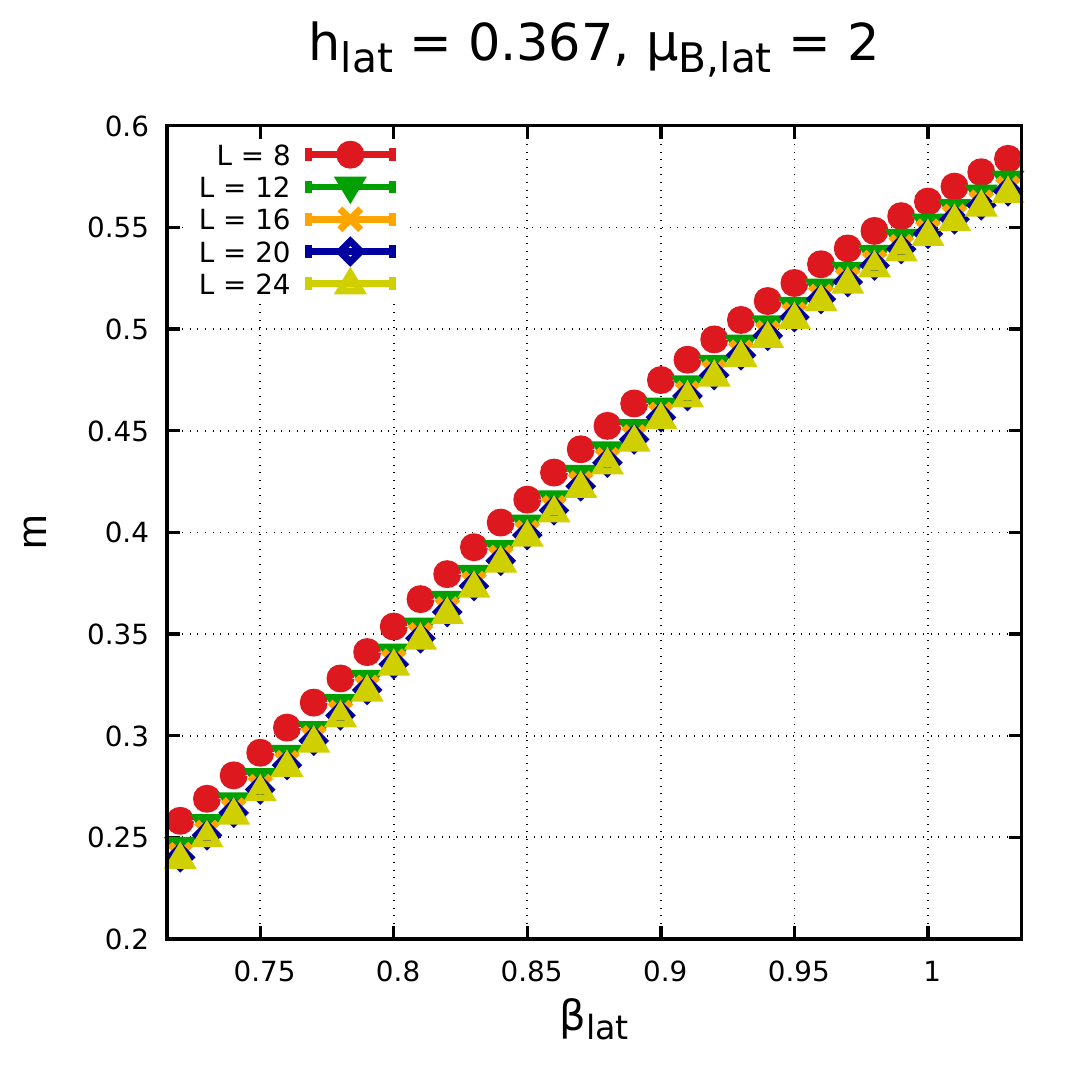}
\hspace*{10mm}
\includegraphics[angle=0,width=.36\linewidth]{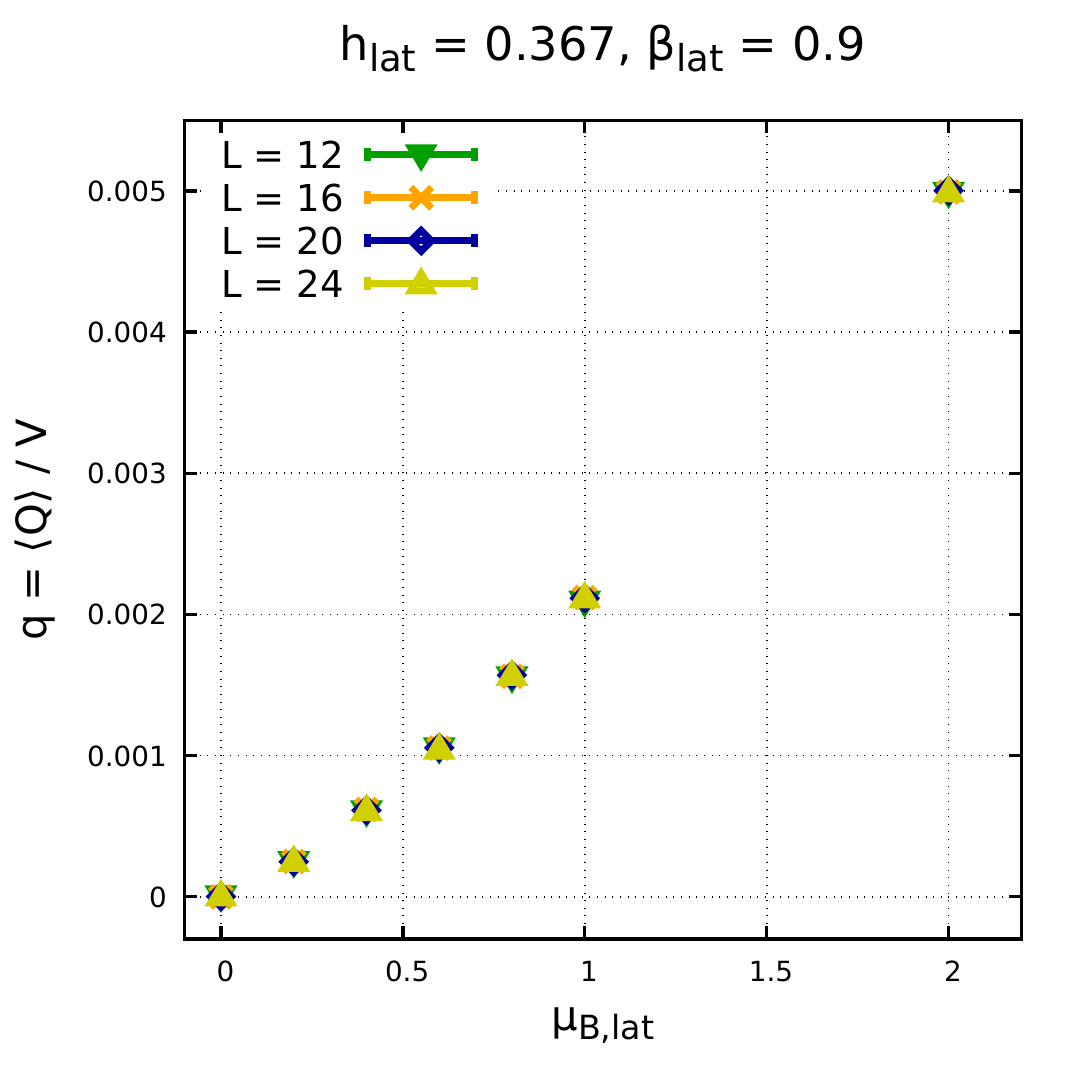} \\
\includegraphics[angle=0,width=.36\linewidth]{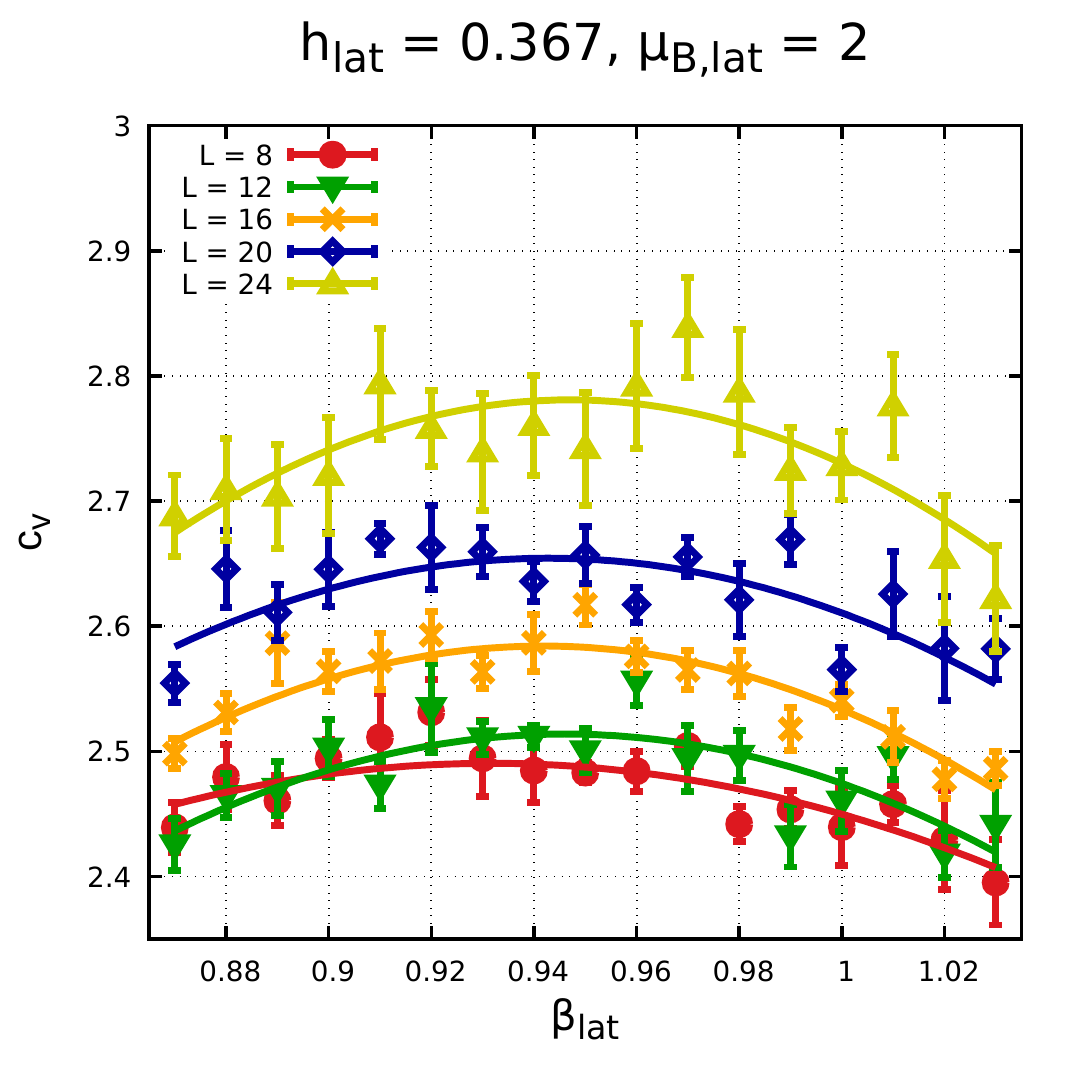}
\hspace*{10mm}
\includegraphics[angle=0,width=.36\linewidth]{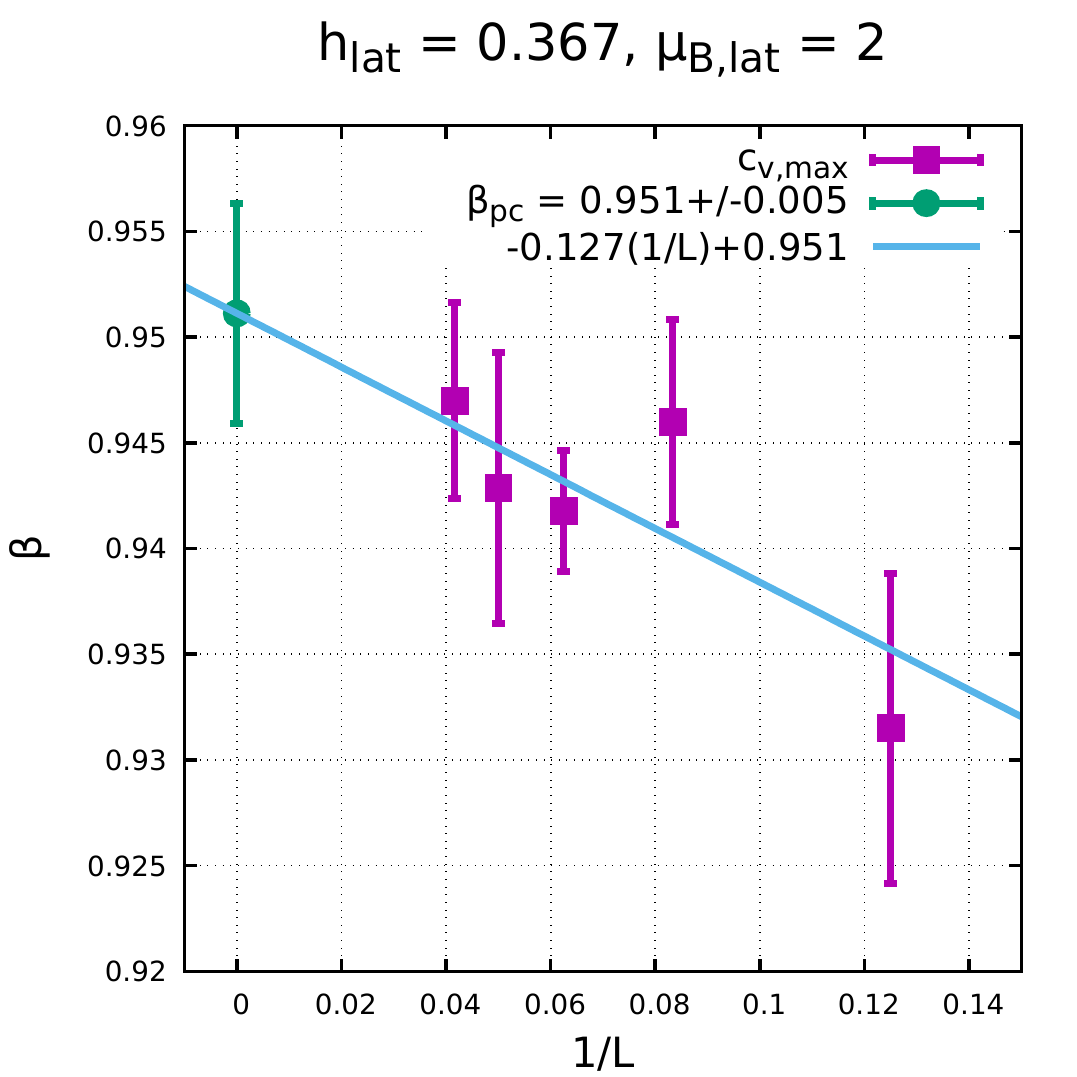} \\
\includegraphics[angle=0,width=.36\linewidth]{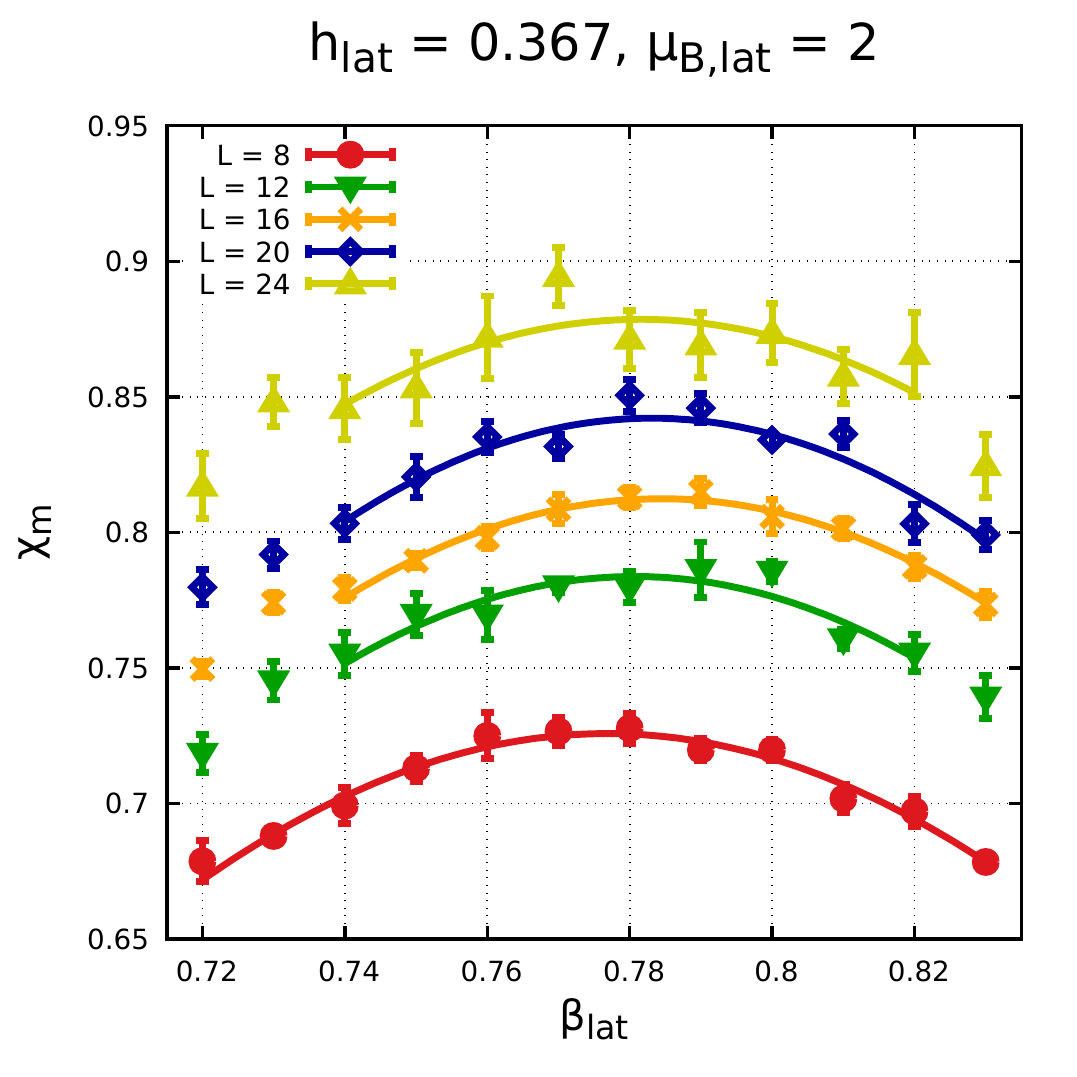}
\hspace*{10mm}
\includegraphics[angle=0,width=.36\linewidth]{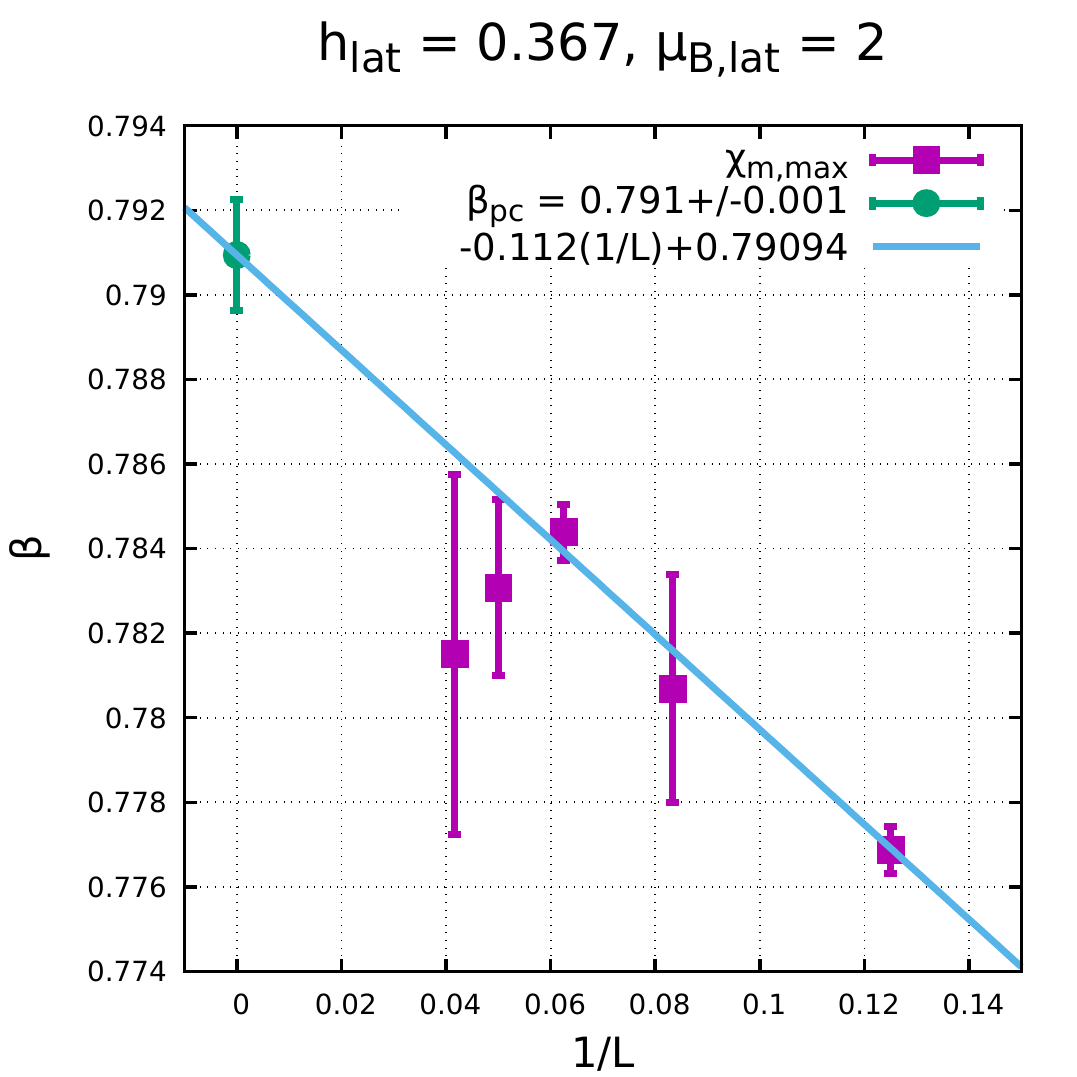}
\end{center}
\vspace*{-6mm}
\caption{Simulation results of the effective theory at $h_{\rm lat} =0.367$,
  compatible with the quark mass $m_{d}$:
  $\tau_{m}$ as an example for the
  auto-correlation time; $\epsilon$, $m$ and $\la Q \ra$ as smooth first
  derivatives of the free energy $F$; $c_{V}$ and $\chi_{m}$ with broad
  maxima, which we identify by Gaussian fits,
  and large-$L$ extrapolations to $\beta_{\rm pc,lat}$.}
\label{h367multiplots}
\vspace*{-12mm}
\end{figure}

Figure \ref{h367multiplots} shows results at $h=0.367$:
the auto-correlation time (top, left) is strongly alleviated
thanks to the crossover; there is no critical slowing down anymore.
The next three plots show the quantities
$\epsilon$, $m$ and $\la Q \ra$ at various values of $L$,
$\mu_{B,{\rm lat}}$ and $\beta_{\rm lat}$. $\epsilon$ depends somewhat
on $\mu_{B,{\rm lat}}$, but hardly on $L$, while for $m$ some
$L$-dependence is visible. These three observables are all
smooth; there is no interval of a steep slope, in contrast to
Figure \ref{h0multiplots}.
This is again consistent with the fact that the second order phase
transition of the chiral limit is now washed out to a crossover.

For the same reason, $c_{V}$ and $\chi_{m}$ do not exhibit actual
peaks but rather smooth maxima, as we see in the lower two plots
in the left column of Figure \ref{h367multiplots}. We identify
these maxima with Gaussian fits,
and estimate their uncertainties with the jackknife method.
The corresponding plots on the right illustrate the
extrapolations to $\beta_{\rm pc,lat}$ in the large-$L$ limit.
\begin{figure}
\vspace*{-5.7cm}
\begin{center}
\hspace*{1.3cm}
\includegraphics[angle=0,width=1.15\linewidth]{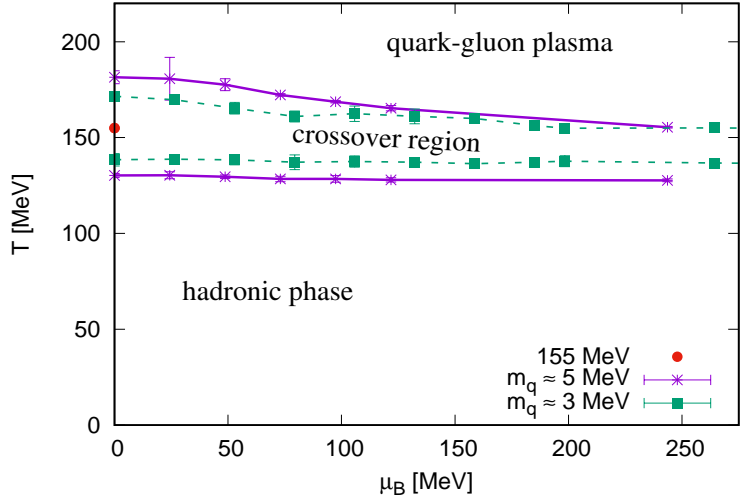}
\end{center}
\vspace*{-2cm}
\caption{Conjectured phase diagram of 2-flavor QCD at
  $m_{\rm q} \approx 5 \ {\rm MeV}$, and at $m_{\rm q} \approx 3 \ {\rm MeV}$.}
\vspace*{-3mm}
\label{massivephasediagram}
\end{figure}

Such extrapolations are performed at each $\mu_{B,{\rm lat}}$-value,
in order to monitor the crossover. There is a significant difference
between extrapolated $\beta_{\rm pc,lat}$-results obtained from $c_{V}$
and from $\chi_{m}$, which is again typical for a crossover.
If we take them as the upper and lower bound of the crossover region,
and convert to physical units as describe in the beginning
of this section, then $T_{\rm pc} \simeq 155 \ {\rm MeV}$ is in the center
of the crossover interval at $\mu_{B}=0$.
Thus we arrive at the phase diagram in Figure \ref{massivephasediagram},
where we include the results obtained at $h=0.14$, $L=24$.
We only see a slight trend for $T_{\rm pc}$ to decrease when $\mu_{B}$
increases, and there is again no indication of a CEP in the range that
we explored, in this case up to $\mu_{B} \approx 250 \ {\rm MeV}$.

\vspace*{-2mm}
\section{Conclusions}
\vspace*{-2mm}

We assume the O(4) model to be in the universality class of 2-flavor
QCD in the chiral limit (O($N$)-models cannot cope with more
than 2 quark flavors). We refer to high temperature, which suggests
a dimensional reduction  to the 3d O(4) model, with a topological
charge $Q$, which corresponds to the baryon number $B$. This effective
theory can be simulated at finite baryon chemical potential $\mu_{B}$,
with a powerful cluster algorithm, and without any sign problem.

In the chiral limit, we monitored the critical line from
$(\mu_{B}, T_{\rm c}) \simeq (0, 132 \ {\rm Mev})$ to
$(309 \ {\rm MeV},$ $ 106 \ {\rm MeV})$. In this range,
$T_{\rm c}(\mu_{B})$ decreases monotonically. No
CEP is found, but there are hints for it to be near the endpoint
of our study, so far.

\begin{figure}
\vspace*{-8mm}
\begin{center}
\includegraphics[angle=0,width=.47\linewidth]{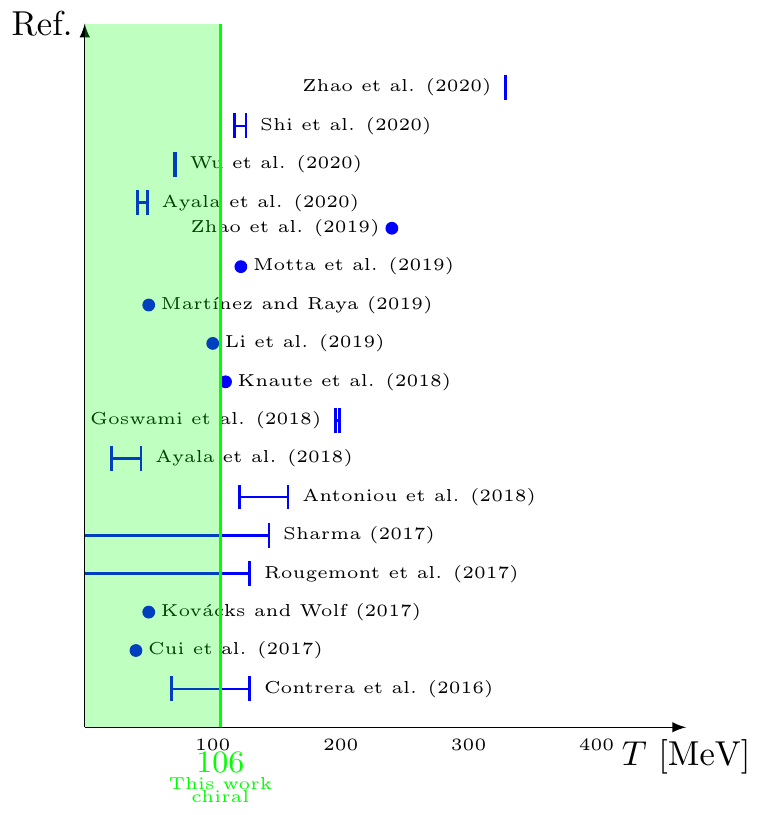}
\hspace*{5mm}  
\includegraphics[angle=0,width=.47\linewidth]{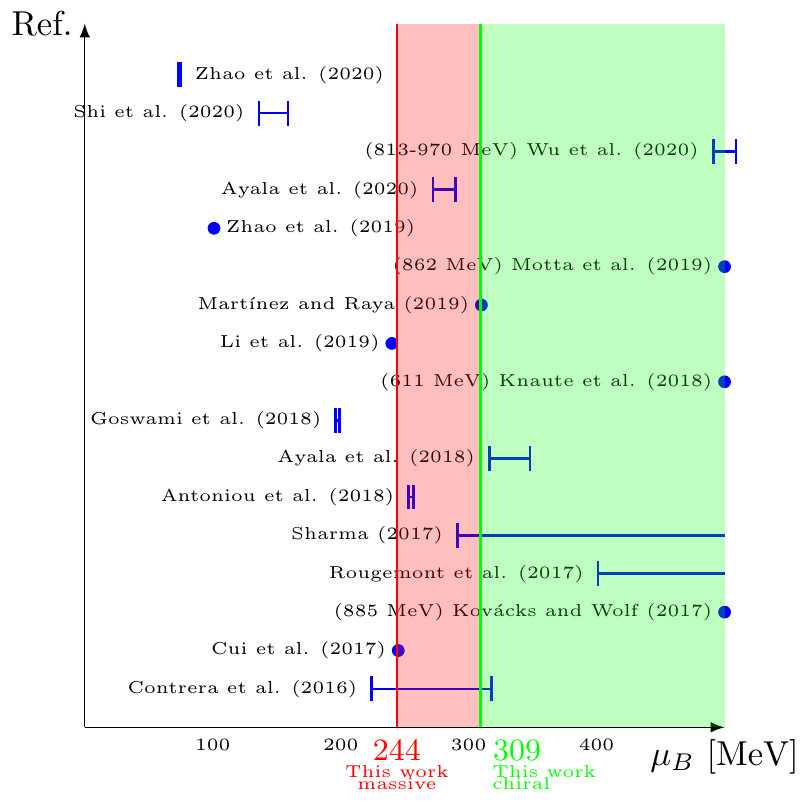}
\end{center}
\vspace*{-5mm}
\caption{The allowed regions of $T_{\rm CEP}$ and $\mu_{B,{\rm CEP}}$
as conjectured here, and in a number of other works.}
\vspace*{-4mm}
\label{CEPlit}
\end{figure}

If we add an external magnetic field, which roughly corresponds to
the light quark masses, $T_{\rm pc}(\mu_{B})$ varies only little
up to $\mu_{B} \approx 250 \ {\rm MeV}$, and again there is no sign
of a CEP. Figure \ref{CEPlit} compares our bounds on $T_{\rm CEP}$ and
$\mu_{B, \rm CEP}$ to a multitude of other conjectures in the literature.\\

\noindent
{\small {\bf Acknowledgments:} WB thanks Uwe-Jens Wiese for instructive
discussions. Arturo Fern\'{a}ndez T\'{e}llez and Miguel \'{A}ngel Nava
Blanco contributed to this project at an early stage \cite{Nava}.
The simulations were carried out on the cluster of ICN-UNAM.
This work was supported by UNAM-DGAPA through PAPIIT project IG100219,
``Exploraci\'{o}n te\'{o}rica y experimental del diagrama de fase de la
cromodin\'{a}mica cu\'{a}ntica'', and by the
Consejo Nacional de Ciencia y Tecnolog\'{\i}a (CONACYT).}


\begin{thebibliography}{99}
\vspace*{-2mm}
  
\bibitem{Ding} H.-T.\ Ding {\it et al.} (HotQCD Collaboration),
  Phys.\ Rev.\ Lett.\ 123 (2019) 062002.

\bibitem{Kotov} A.Yu.\ Kotov, M.P.\ Lombardo and A.\ Trunin,
arXiv:2105.09842 [hep-lat].

\bibitem{BazavovBhattacharya} A.\ Bazavov {\it et al.}\ (HotQCD Collaboration),
  Phys.\ Lett.\ B 795 (2019) 15.
T.\ Bhattacharya {\it et al.}\ (HotQCD Collaboration),
Phys.\ Rev.\ Lett.\ 113 (2014) 082001.

\bibitem{Philippe} P.\ de Forcrand,
PoS LAT2009 (2009) 010.

\bibitem{chiralPT} R.D.\ Pisarski and F.\ Wilczek,
Phys.\ Rev.\ D 29 (1984) 338.
F.\ Wilczek,
Int.\ J.\ Mod.\ Phys.\ A 7 (1992) 3911.
K.\ Rajagopal and F.\ Wilczek,
Nucl.\ Phys.\ B 399 (1993) 395.
For a recent review, see A.Yu.\ Kotov, M.P.\ Lombardo and A.\ Trunin,
Symmetry 13 (2021) 1833.
    
\bibitem{Skyrme} T.H.R.\ Skyrme,
Proc.\ Roy.\ Soc.\ Lond.\ A 260 (1961) 127;
Nucl.\ Phys.\ (1962) 31 556.
G.S.\ Adkins, C.R.\ Nappi and E.\ Witten,
Nucl.\ Phys.\ B 228 (1983) 552.
I.\ Zahed and G.E.\ Brown,
Phys.\ Rept.\ 142 (1986) 1.
 
\bibitem{Murakami} J.\ Murakami,
 Proc.\ Amer.\ Math.\ Soc.\ 140 (2012) 3289.

\bibitem{Edgar} E.\ L\'{o}pez-Contreras,
B.Sc.\ thesis, Universidad Nacional Aut\'{o}noma de M\'{e}xico, 2021.

\bibitem{Oevers} M.\ Oevers, Diploma thesis,
  Universit\"{a}t Bielefeld, 1996.
J.\ Engels, L.\ Fromme and M.\ Seniuch,
Nucl.\ Phys.\ B 675 (2003) 533.

\bibitem{Wang} J.-S.\ Wang,
Physica A 161 (1989) 249.

\bibitem{JAGH} J.A.\ Garc\'{\i}a-Hern\'{a}ndez,
B.Sc.\ thesis, Universidad Nacional Aut\'{o}noma de M\'{e}xico, 2020.

\bibitem{ForKas}  C.M.\ Fortuin and P.W.\ Kasteleyn,
Physica 57 (1972) 536.
I.\ Dimitrovi\'{c}, P.\ Hasenfratz, J.\ Nager and F.\ Niedermayer,
Nucl.\ Phys.\ B 350 (1991) 893.

\bibitem{Nava} M.A.\ Nava Blanco, W.\ Bietenholz and
  A.\ Fern\'{a}ndez T\'{e}llez,
  J.\ Phys.\ Conf.\ Ser.\ 912 (2017) 012048.
  M.A.\ Nava Blanco,
M.Sc.\ thesis, Benem\'{e}rita Universidad Aut\'{o}noma de Puebla, 2019.
  
\end{thebibliography}
\end{document}